# Self-assembled physical unclonable function labels based on plasmonic coupling

*Mihir Dass,[1] Lena Raab,[1] Christoph Pauer,[1] Christoph Sikeler,[1] Larissa Heinze,[1] Joe Tavacoli,[1] Irina V. Martynenko,[1] Ulrich Rührmair*,[2,3] Gregor Posnjak*,[1] Tim Liedl*[1]*

[1]Faculty of Physics and Center for NanoScience (CeNS)
Ludwig-Maximilian-University Munich, 80539 Munchen, Germany

[2]Electrical and Computer Engineering Department,
University of Connecticut, Storrs, CT 06249

[3]Institute for Computer Science,
Ludwig-Maximilian-University Munich, 80538 München, Germany

Corresponding authors:

Ulrich Rührmair - ruehrmair@ilo.de
Gregor Posnjak - gregor.posnjak@lmu.de,
Tim Liedl - tim.liedl@physik.lmu.de

## Abstract

Counterfeiting threatens human health, social equity, national security and global and local economies. Hardware-based cryptography that exploits physical unclonable functions (PUFs) provides the means for secure identification and authentication of products. While optical PUFs are among the hardest to replicate, they suffer from low encoding capacity and often complex and expensive read-out. Here we report PUF labels with nanoscale features and optical responses that arise from the guided self-assembly of plasmonic nanoparticles. Nanosphere lithography combined with DNA origami placement are used to create tightly packed randomised nanoparticle assemblies. Nanoscale variations within these assemblies define the scattering colour of the individual spots that are arranged in a hexagonal lattice with spacing down to the optical resolution limit. Due to the nanoscale dimensions, the intrinsic randomness of the particle assemblies and their resulting optical responses, our PUFs are virtually impossible to replicate while they can be read-out with economical 3D-printed hardware.

Combating counterfeiting of goods is a challenge for our society, increasingly so in a globalised world where supply chains become longer and their transparency decreases. Concurrently, the market for sensitive goods, such as medicine, has also increased, further exemplifying the need for anti-counterfeit measures. For example, between 72,000 and 169,000 children die of treatable diseases like pneumonia annually due to counterfeited medicine.[1] The estimated economic losses from counterfeiting are in trillions of dollars.[2] Consequently, increasingly complex and effective authentication tools are being developed. One class of anti-counterfeit measures are *physical unclonable functions* (PUFs), physical objects that provide a unique fingerprint as a result of a stochastic process. Their evaluation is straightforward in the forward direction but infeasible to compute in the reverse one without additional information. Ideally, they are impossible to replicate physically.[3]

As such, a wide range of materials and readout strategies have been shown,[2,4] ranging from carbon nanotubes with electrical current read-out[5], to carbon dots exhibiting fluorescence patterns[6] and block copolymers questioned via Raman scattering[7]. Optical PUFs, that rely on light-matter interactions to produce the fingerprint, are categorised as 'strong' PUFs for the inherent difficulty for an attacker to replicate them.[4] Strong PUFs feature an exponential increase in possible outputs with respect to some system parameter. Among optical elements, plasmonic particles are excellent candidates for programming light-matter interactions because they interact strongly with light. The optical response of plasmonic particles is the result of intrinsic properties like material, shape, size and dielectric environment.[8,9] Additionally, extrinsic factors such as the proximity to other particles can strongly affect their optical behaviour due to near-field mediated plasmonic coupling.[10,11] Combining variations in these intrinsic and extrinsic properties provides the experimenter with a versatile tool box from which to engineer a unique optical response. Indeed, such nanoparticles can interact with light in the visible regime with wavelengths between 400 nm-700 nm, even when their dimensions are sub-100 nm. As a result, several groups have characterised their suitability for building PUFs.[12–14] Practically, these works often suffer from low particle density or low variation in the optical response of the PUF, both of which compromise the security by limiting the complexity of the PUF.

In the current work, we create *PartiPUFs* (Particle PUFs) by combining plasmonic nanoparticle[15] placement (NPP) with DNA Origami Placement (DOP)[16,17] to hexagonally pattern nanoparticle assemblies at a desired surface density using nanosphere lithography (NSL).[18–21] Capture of the assemblies is mediated by first laying down hexagonal arrays of DNA origami *nanodiscs* that capture different combinations of nanoparticles. Plasmonic coupling between the nanoparticles gives rise to new colours absent at the discrete particle level, which can be readily detected using Dark Field microscopy (DFM).[22] Control over the DNA origami design, NSL parameters and the combination and concentration of nanoparticles allows us to tailor the optical response and pack the surface with a high information density. The resulting colourful plasmonic labels allow for the generation of a large number of challenge-response pairs, instrumental in the security of the PUF. Our PUFs demonstrate a unique and broad variation in their optical response, which we characterize by the implementation of a hue analysis. We also fabricated an economical and portable DFM and used it to successfully image our PUFs, showcasing the suitability of our technology for practical applications.

**Fabrication of PartiPUFs**

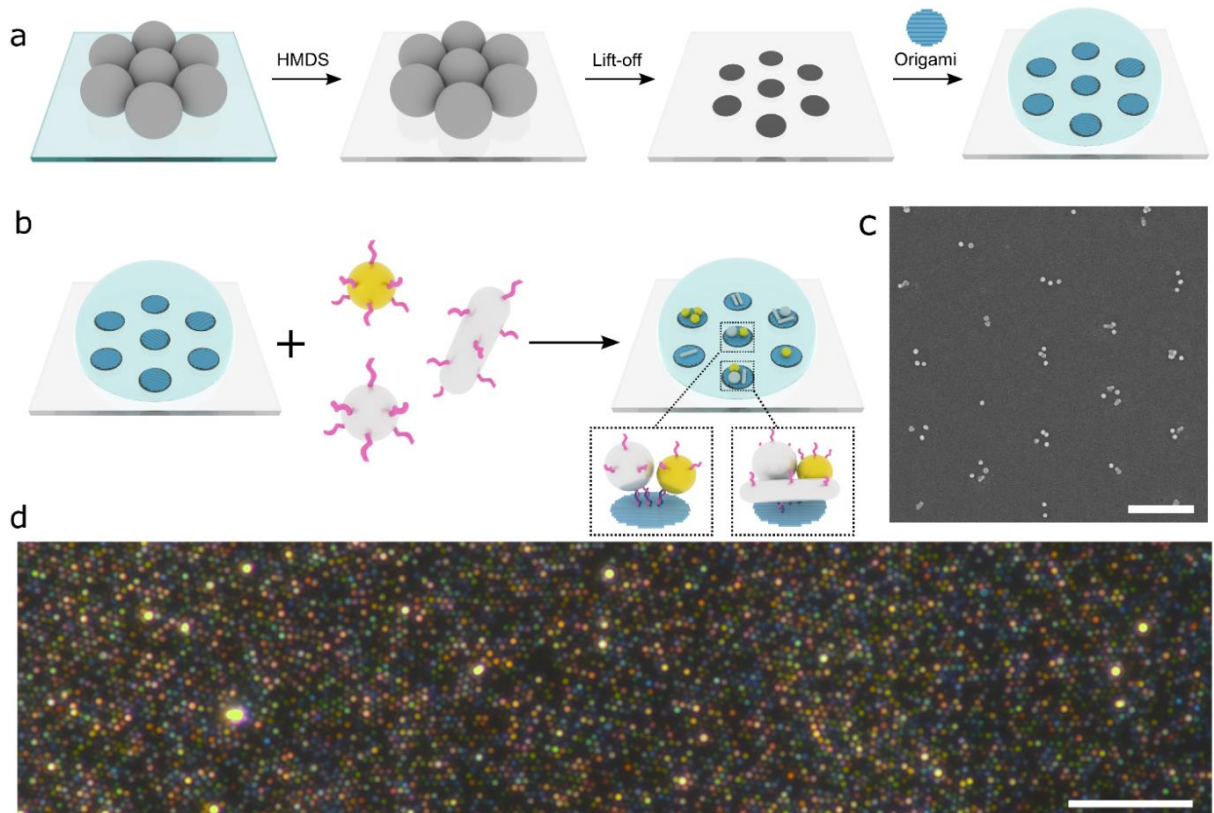

**Figure 1.** Schematic and fabrication of PartiPUFs. a) NSL procedure involving Polystyrene nanosphere deposition, surface passivation using HMDS, lift-off and DOP using DNA origami. b) NPP performed using various functionalised nanoparticles. d) SEM image after NPP. d) DFM image of the resultant PUF. Scale bars, (c) 500 nm and (d) 10 μm.

Figure 1a illustrates the fabrication process of the PartiPUF. Glass coverslips are nanopatterned using NSL to create *placement sites* for the DOP. DOP is then performed to place DNA origami nanodiscs (~ 100 nm x 85 nm) on the glass surface. Each nanodisc contains *binding sites* made of poly-Adenine ssDNA extensions (10 nt long, "$A_{10}$") extended from both faces in a filled circular arrangement (Fig. S1). The desired mix of nanoparticles functionalised with poly-Thymine ssDNA (9:1 mix of $T_8$:$T_{19}$) is then added, where the DNA origami structures serve to capture the nanoparticles through strand hybridisation (Fig. 1b). The result is a stochastic assembly of nanoparticles at each placement site (Fig. 1c) to produce a dense packing of coloured spots when viewed under a DFM (Fig. 1d). A variety of plasmonic particles – gold nanospheres (AuNS), gold nanorods (AuNR), gold/silver core/shell nanospheres (AgNS) and nanorods (AgNR) – are used to create a broad variation in the optical response.[23,24] Plasmonic coupling between nanoparticles placed sufficiently close together (1-10 nm) gives rise to new colours in the assemblies, further increasing the output complexity.

We can direct the appearance of our PUF in two main ways. Firstly, we can adapt the pitch size between the individual scattering spots by adjusting the size of the nanospheres employed during NSL (Fig. S2). The nanosphere diameter defines both the size of the placement site and the spacing between placement sites, which is approximately equal to the diameter of the spheres used.[17] We performed NSL using polystyrene nanospheres with diameters of 400, 600 and 1000 nm. Controlling the spacing between the spots is crucial because it allows to increase the density of our spots to a critical point just above the diffraction limit, below which the colours would 'bleed 'into each other making optical analysis infeasible. The DNA origami structures used for subsequent deposition feature a binding site containing 42 $A_{10}$ extensions (21 from each face, distributed in a circular pattern with a diameter of ~ 60 nm) to capture an undefined number of nanoparticles.

Secondly, we can adapt the colour distribution of our PUF by varying the material, form and population composition of the plasmonic particles trapped at each DNA origami. Specifically, four different types of nanoparticles were utilised in our experiments: 30 nm AuNS, 50 nm AgNS, AuNR with length = 60 nm and width = 12 nm and AgNR with length = 75 nm and diameter = 25 nm (Supplementary note S1). The silver-coating enhances the scattering of the nanoparticles and provides variation in the scattering colour of the individual nanoparticle species.[25] The silver-coating also pushes the LSPR wavelength of the nanorods into the visible, thus ensuring that all nanoparticle species scatter in the visible regime. Next to the controlled spacing between the assembly sites, the NSL method provides a cheap, fast and parallel way to program the size of the placement spots and thus, with the help of DNA origami patches, also ensures tight spacing between the individual nanoparticles.[17,18] The result is a random distribution of scattering colours due to the type of particles and the plasmonic interactions between them on each placement site.

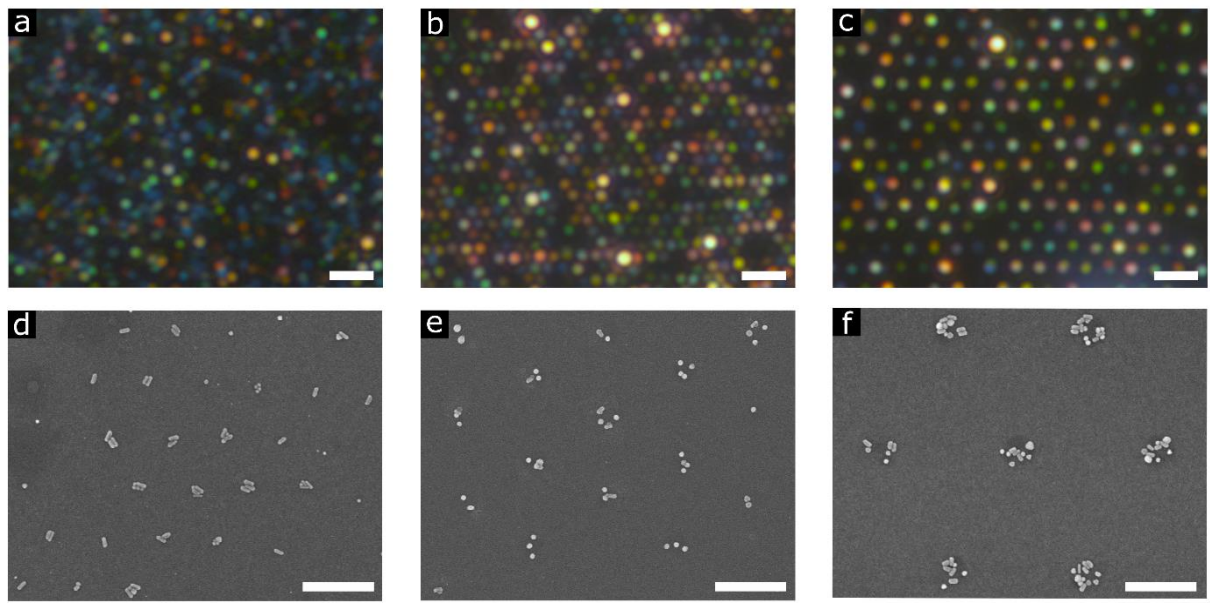

**Figure 2.** PUFs fabricated with varying inter-site spacing. a-c) DFM images (full field-of-view in Fig. S5-S7) and d-f) SEM images of PartiPUFs made with a,d) 400 nm, b,e) 600 nm and c,f) 1 μm spacing (uncropped images: Fig. S8-S10). Scale bars, (a-c) 2 μm, (d-f) 500 nm.

We fabricated PUFs with varying site spacing and a wide range of colours scattering from the individual placement sites (Fig. 2a-c). The strong scattering of the nanoparticles enabled fast acquisition times in DFM, which makes it a suitable characterisation method for our PUFs.[26] The SEM images show the stochastic nature of the assembly of particles at the individual placement sites (Fig. 2d-f). For the smallest spheres used in our experiments (400 nm, Fig. 2a,d), we observed small groups of densely placed particles on each contact spot. Larger spheres (600 nm and 1 μm, Fig. 2b,c,e,f) resulted in larger contact areas, often leading to the binding of more than a single DNA origami structure. Accordingly, we observed more particles bound per spot with discernible gaps between the particles or groups of particles. However, the number of DNA origami structures per placement site can be further tuned with parameters such as the $Mg^{2+}$ or the DNA origami concentration or the incubation time.[16] In the case of our nanodisc DNA origami structure, a nanosphere with a diameter of 350 nm results in single-origami placement (Fig. S11). Using larger spheres leads to multiple structures being placed on each placement site (Fig. 2e,f and Fig. S12). In such cases, we optimised the DNA origami concentration during DOP to obtain the desired number of DNA origami structures on the placement sites. High DNA origami concentrations (450 pM) led to complete saturation of the placement sites (Fig. S12d).

**PartiPUF characterization and Hue analysis**

We analysed the DFM images of PartiPUFs using an ad hoc-written image analysis algorithm (Supplementary note S2). For our analysis we chose the Hue-Saturation-Brightness (HSB) scale as it uses a single value - the 'hue 'as a proxy for the wavelength - to identify colour. Hue values of the individual scattering spots thus quantitatively characterise the complexity of the scattering response of our PUFs (Fig. S13). We fabricated PartiPUFs with different nanoparticle species and concentrations (Fig. 3a-f) and obtained a wide range of hues and hue distributions (Fig. 3g,h). The optical density (O.D.) of the nanoparticle suspensions was used to measure their concentration. Higher concentrations of nanoparticles (usually O.D. > 1) resulted in hue distributions skewed towards lower, i.e. redder, values, even when using multiple nanoparticle species (Fig. 3g). PUFs made using only AgNSs illustrate the effect of variations in the particle concentration (Fig. 3a,d). Using higher concentrations of AgNS (O.D. ~ 0.75) (Fig. 3d) resulted in a relatively narrow hue distribution, with a higher frequency of hues at lower values in the range of $20 - 40$ (Fig. 3h). Decreasing the AgNS concentration slightly (O.D. ~ 0.5) led to a bimodal distribution (Fig. 3g) with a peak appearing at hue ~ 100, similar to that of individual AgNS deposited on a surface (Fig. S21). The secondary maximum, around hue ~ 40, is consistent with aggregates giving lower hue values. An elevated concentration (O.D. ~ 1.5) of a mixture of AuNSs and AuNRs (Fig. 3b) resulted in hues with a peak of the distribution at hue ~ 25. Lower concentrations (O.D. ~ 0.5) of the same particle mix (Fig. 3e) shifted the hue distribution towards higher hue values. High concentrations (O.D. ~ 2.7) of a mixture of AuNSs, AgNRs and AgNSs (Fig. 3c) resulted in a narrow hue distribution, similar to that obtained for particle mixes containing AuNSs and AuNRs. Lower concentrations (O.D. ≤ 0.5) of a mixture of AuNSs, AgNRs and AgNSs (Fig. 3f) resulted in the broadest range of hues (Fig. 3h). Hues ranging from 6 to 166 were achieved, covering red, orange, yellow, green and blue. We could also vary the hue distribution of PUFs by controlling the number of particles attaching to each DNA origami structure with the number of poly-A particle capture extensions (Supplementary note S3).

Under ideal conditions, i.e. each hue value would be unmistakably recognised during the imaging process, the number of combinations in a typical image of our PUFs can be given by

$$\text{Number of combinations} = (\textit{Number of Hues})^{\textit{number of scattering spots}}$$

For the uncropped image (Fig. S20) of the PUF made with 1 µm spacing in Fig. 3f, with a field-of-view of 138 µm x 69 µm at 100x magnification, the number of combinations would be $152^{7,753} \approx 10^{16,915}$. Using the same imaging parameters for the uncropped image (Fig. S6) of a PUF made with 600 nm spacing (Fig. 1d), this number would increase to $160^{16,640} \approx 10^{36,676}$. The exponential increase in the number of combinations with increasing number of scattering spots places these objects in the category of strong PUFs.[27]

We tested the dependence of the hue value returned by our image analysis algorithm with respect to imaging conditions such as lamp intensities and exposure times and found minimal variation (see Table S5). In practical usage, binning similar hues to account for lighting differences would minimise the effect of this variation and make the imaging more robust. When using hue binning, the number of combinations can be given by

$$\text{Number of combinations} = (\textit{Number of Hue bins})^{\textit{number of scattering spots}}$$

We recorded a hue variation of ± 2 from a hue = 20 measured under our standard imaging conditions (Table S5). Using a conservative hue binning value of 5 would result in $10^{11,773}$ possible combinations for the 1 µm spacing sample (Fig. S20) and $10^{25,268}$ for the sample with 600 nm spacing (Fig. S6).

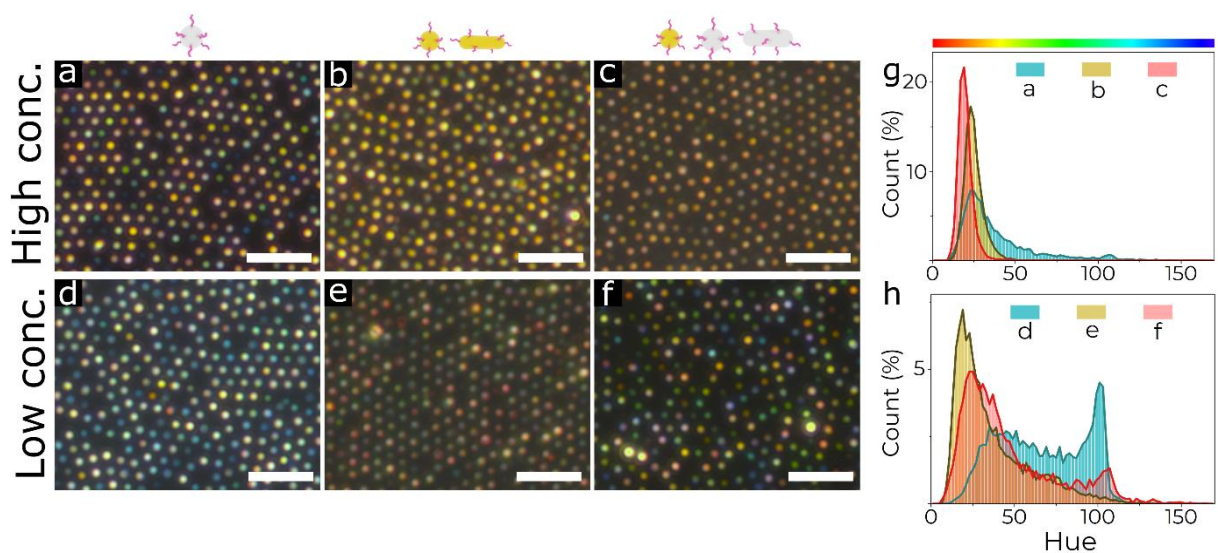

**Figure 3.** Hue analysis of PUFs. a-f) DFM images (full field-of-view in Fig. S15-20) of PUFs created while varying nanoparticle species and concentration. The nanoparticle species used are shown at the top of each column. g,h) Hue distribution of PUFs fabricated using different particle species at (g) higher and (h) lower concentrations, calculated from the full field-of-view images. Number of spots analysed for the histograms, n = 7700. For placement conditions, see Table S4. A hue scale corresponding to the x-axis is shown above the plots. Scale bars, 5 µm.

**Enhancing security via anisotropic optical response**

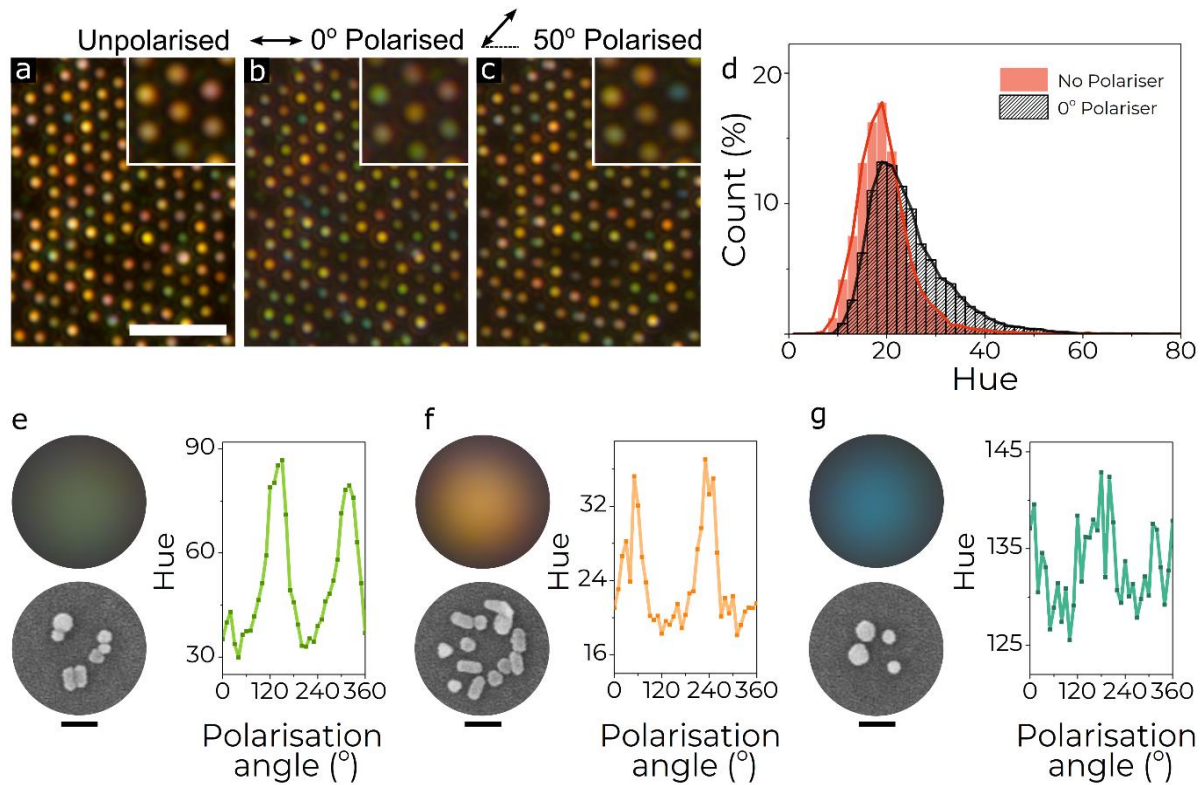

**Figure 4.** Polarisation-dependent behaviour of the PartiPUFs. a-c) DFM images of the same area at different polariser orientations. Insets show a magnified view of the same seven spots. Double-sided arrows depict the orientation of the polariser with respect to the image. d) Variation in hue distribution of PUFs in the presence of a polariser. e-g) Unpolarised DFM image (top), SEM image (bottom) and the polarisation-depended hue responses of three individual particle assemblies. Scale bars, (a-c) 5 μm, (e-g) 100 nm.

Asymmetric plasmonic particles, such as nanorods, possess longitudinal and transverse plasmon modes along their long and short axes respectively.[28] Therefore, they exhibit an angle-dependent scattering response when excited by polarised light.[29] This results in a polarisation-angle-dependent change in the scattering colour of the nanorods when viewed under DFM.[30,31] Additionally, plasmonic coupling can give rise to collective modes in assemblies of particles in close proximity, which also display such polarization-sensitive responses.[32] The nanoparticle assemblies in our PUFs display similar behaviour, resulting in a changing hue when imaged while varying the polarisation angle of the incident light (Fig. 4a-c). Additionally, the distribution of the hue values changes when imaged under static polarised illumination (Fig. 4d). We also observed that complex assemblies composed of multiple particles exhibit a variety of hue responses under changing polarisation illumination (Fig. 4e-g). SEM images reveal that the nanoparticle assemblies often contain multiple nanorods and spheres at varying distances from each other. In all cases, we see an angle-dependent hue response. These fall into two classes, those that show well-defined peaks (Fig. 4e,f) and those that show a more 'noisy' hue response (Fig. 4g). The former suggests that collective plasmonic modes of such assemblies possess a unique long-axis. However, this is not intuitively evident from the arrangement of the particles within the assemblies. The polarisation-dependent hue response can enhance the security of the PUF by challenging the authenticator to reproduce the hue response at any arbitrary angle along with the unpolarised response. Accordingly, an attacker would need the ability to clone both responses to fool the authenticator.

**Practical implementation of our PartiPUFs**

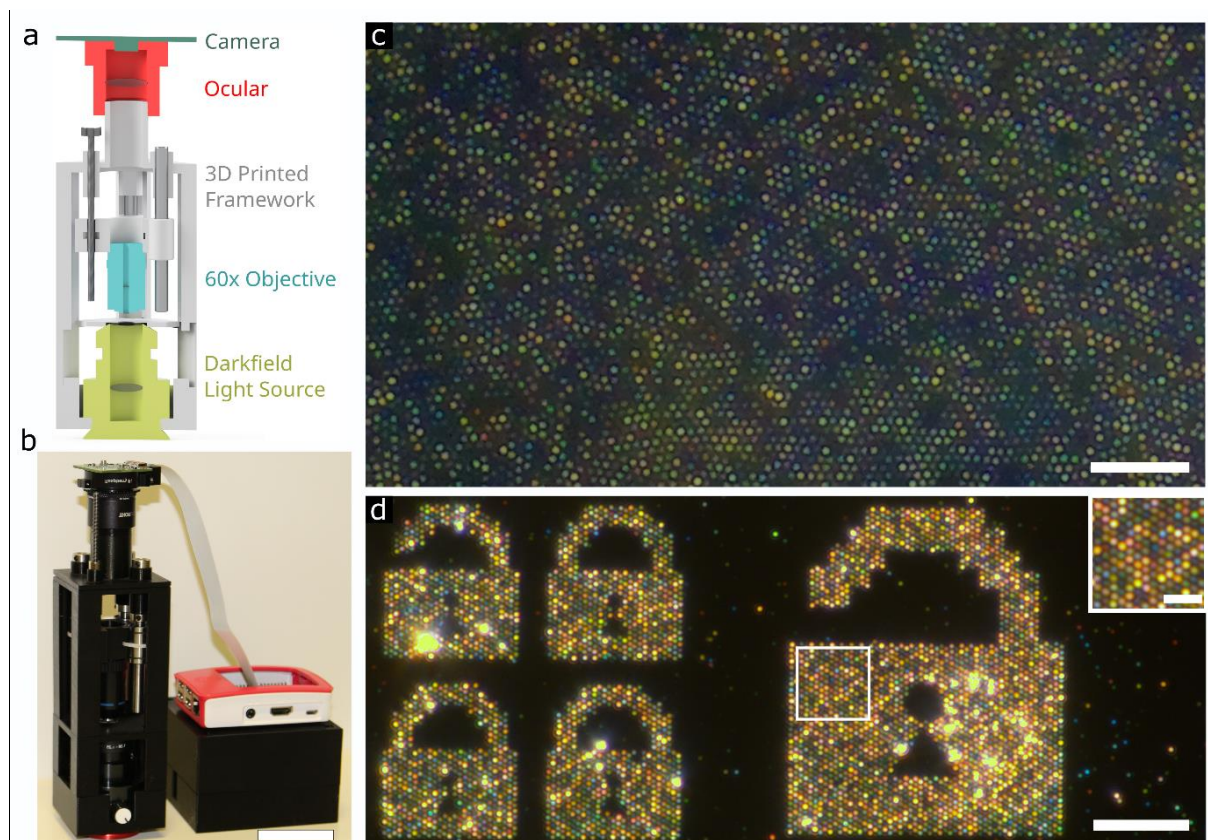

**Figure 5.** Practical implementation of the PartiPUFs. a) Schematic and b) realisation of a 3D-printed Dark Field microscope (3DFM). c) DFM image of the PUF in Fig. 3d (Fig. S18) taken with the 3DFM at 60x magnification. d) A PUF assembled using AuNS and AgNS on a 'lock' pattern with 600 nm spacing made by e-beam lithography. Inset: a magnified view showing the hexagonal array of scattering spots. Scale bars, (b) 5 cm, (c,d) 10 μm and inset 2.5 μm.

Unlike other high-security implementations of PUFs which require sophisticated and costly instruments to verify the tag, our PartiPUFs can be questioned with an inexpensive optical setup. To demonstrate this, we designed a simplified 3-D printed DFM, which we then used to image our PUFs. The 3-D printed DFM consists of a dark-field condenser, an achromatic 60x objective, an ocular lens and a 12-megapixel Raspberry Pi camera module (Fig. 5a,b). We used a 3D-printed framework to align the optical components in the horizontal direction. We can adjust the height of the dark-field condenser to ensure contact between the condenser, the immersion oil and the sample. The objective is connected to three steel rods, which are used to move it along the z-direction to allow for focusing. Using this setup, we imaged the PUF in Fig. 3d and were able to resolve the scattering spots and their hues (Fig. 5c). We estimate that on a commercial scale a PartiPUF testing machine/read-out machine could be produced from components costing on the order of ~200 € (Supplementary note S6).

Our process is also compatible with other lithography methods such as e-beam or nanoimprint lithography which can form consistent and spatially variable placement pattern. This can be used to encode additional safety features or easily recognisable markers to locate the read-out region of the tag. To demonstrate this, we performed DOP and NPP on substrates patterned with e-beam lithography.[33,34] Figure 5d shows the DFM image of 'locks' made up of a hexagonal array of placement sites. Additionally, DOP has been shown to be compatible with techniques like nanoimprint lithography.[16] This would allow for printing of predefined patterns like product barcodes on which NPP can be performed, combining product data and authentication.

**Conclusions**

We have shown that self-assembly of plasmonic nanoparticles, driven by DNA origami, can be used to create nanoscale PUF labels with high information density. Our approach combines a unique optical response on the micron scale with information encoding on the nanoscale, which is accessible with a relatively simple and inexpensive optical microscope. To fabricate our PUF labels we used an approach that does not demand any sophisticated instrumentation and intrinsically generates unique objects because of the stochastic nature of their assembly. While each PUF labels is unique, we have shown we can control the range and distribution of accessible hue values of our PUFs. Because the optical response of each scattering site depends on the size, shape and composition of multiple nanoparticles, an attacker would have to first characterise a specific PUF with sophisticated instrumentation and then replicate it on the nanometre scale using several materials, which is infeasible even using top-end lithography manufacturing. We also show seamless integration of our PartiPUF with a cheap read-out tool in the form of a 3D-printed DFM, which significantly bridges the gap between experimental technology and its application.

**ACKNOWLEDGMENTS**

We thank Philipp Altpeter and Christian Obermayer for clean room assistance and Susanne Kempter for assistance with TEM. M.D. and this work was funded by the Federal Ministry of Education and Research (BMBF) and the Free State of Bavaria under the Excellence Strategy of the Federal Government and the Länder through the ONE MUNICH Project Munich Multiscale Biofabrication and the project Enabling Quantum Communication and Imaging Applications. J.T. is funded under DFG TA 1375/2-1 and C.P. through the DFG SFB1032 "Nanoagents," Project A6. C.S., I.M., G.P., and T.L. acknowledge funding from the ERC consolidator grant "DNA Funs" (Project ID: 818635).

**AUTHOR CONTRIBUTIONS STATEMENT**

T.L., U.R., G.P. and M.D. designed this study. M.D. and T.L. designed the DNA origami structures. M.D., L.R. and L.H. assembled and purified the DNA origami structures, synthesised and functionalised the nanoparticles and performed the placement experiments and their characterisation. C.P. wrote the ad-hoc image analysis script. C.P. and J.T. performed the image analysis. G.P. and C.S. helped with dark field imaging. C.S. designed and fabricated the 3D-printed DFM and used it for imaging PUFs. I.M. designed the 'lock' patterns and performed e-beam lithography. M.D., G.P. and T.L. wrote the manuscript with input from all authors.

**COMPETING INTERESTS STATEMENT**

The authors declare no competing financial interest.

**REFERENCES**

1. OECD/EUIPO (2020), Trade in Counterfeit Pharmaceutical Products, Illicit Trade, OECD Publishing, Paris, https://doi.org/10.1787/a7c7e054-en.

2. Arppe, R. & Sørensen, T. J. Physical unclonable functions generated through chemical methods for anti-counterfeiting. *Nat Rev Chem* **1**, 1–13 (2017).

3. Pappu, R., Recht, B., Taylor, J. & Gershenfeld, N. Physical One-Way Functions. *Science* **297**, 2026–2030 (2002).

4. Gao, Y., Al-Sarawi, S. F. & Abbott, D. Physical unclonable functions. *Nat Electron* **3**, 81–91 (2020).

5. Hu, Z. *et al.* Physically unclonable cryptographic primitives using self-assembled carbon nanotubes. *Nature Nanotech* **11**, 559–565 (2016).

6. Zhang, J. *et al.* An all-in-one nanoprinting approach for the synthesis of a nanofilm library for unclonable anti-counterfeiting applications. *Nat. Nanotechnol.* 1–9 (2023) doi:10.1038/s41565-023-01405-3.

7. Kim, J. H. *et al.* Nanoscale physical unclonable function labels based on block copolymer self-assembly. *Nat Electron* **5**, 433–442 (2022).

8. Wang, L., Hasanzadeh Kafshgari, M. & Meunier, M. Optical Properties and Applications of Plasmonic-Metal Nanoparticles. *Advanced Functional Materials* **30**, 2005400 (2020).

9. Martens, K. *et al.* Onset of Chirality in Plasmonic Meta-Molecules and Dielectric Coupling. *ACS Nano* **16**, 16143–16149 (2022).

10. Rechberger, W. *et al.* Optical properties of two interacting gold nanoparticles. *Optics Communications* **220**, 137–141 (2003).

11. Jain, P. K., Eustis, S. & El-Sayed, M. A. Plasmon Coupling in Nanorod Assemblies: Optical Absorption, Discrete Dipole Approximation Simulation, and Exciton-Coupling Model. *J. Phys. Chem. B* **110**, 18243–18253 (2006).

12. Smith, A. F., Patton, P. & Skrabalak, S. E. Plasmonic Nanoparticles as a Physically Unclonable Function for Responsive Anti-Counterfeit Nanofingerprints. *Advanced Functional Materials* **26**, 1315–1321 (2016).

13. Lu, Y. *et al.* Plasmonic Physical Unclonable Function Labels Based on Tricolored Silver Nanoparticles: Implications for Anticounterfeiting Applications. *ACS Appl. Nano Mater.* **5**, 9298–9305 (2022).

14. Li, Q. *et al.* Physical Unclonable Anticounterfeiting Electrodes Enabled by Spontaneously Formed Plasmonic Core–Shell Nanoparticles for Traceable Electronics. *Advanced Functional Materials* **31**, 2010537 (2021).

15. Rothemund, P. W. K. Folding DNA to create nanoscale shapes and patterns. *Nature* **440**, 297–302 (2006).

16. Gopinath, A. & Rothemund, P. W. K. Optimized Assembly and Covalent Coupling of Single-Molecule DNA Origami Nanoarrays. *ACS Nano* **8**, 12030–12040 (2014).

17. Shetty, R. M., Brady, S. R., Rothemund, P. W. K., Hariadi, R. F. & Gopinath, A. Bench-Top Fabrication of Single-Molecule Nanoarrays by DNA Origami Placement. *ACS Nano* **15**, 11441–11450 (2021).

18. Deckman, H. W. & Dunsmuir, J. H. Natural lithography. *Applied Physics Letters* **41**, 377–379 (1982).

19. Haynes, C. L. & Van Duyne, R. P. Nanosphere Lithography: A Versatile Nanofabrication Tool for Studies of Size-Dependent Nanoparticle Optics. *J. Phys. Chem. B* **105**, 5599–5611 (2001).

20. Qiu, T. *et al.* Nanosphere Lithography: A Versatile Approach to Develop Transparent Conductive Films for Optoelectronic Applications. *Advanced Materials* **34**, 2103842 (2022).

21. Parchine, M., McGrath, J., Bardosova, M. & Pemble, M. E. Large Area 2D and 3D Colloidal Photonic Crystals Fabricated by a Roll-to-Roll Langmuir–Blodgett Method. *Langmuir* **32**, 5862–5869 (2016).

22. Tian, X., Zhou, Y., Thota, S., Zou, S. & Zhao, J. Plasmonic Coupling in Single Silver Nanosphere Assemblies by Polarization-Dependent Dark-Field Scattering Spectroscopy. *J. Phys. Chem. C* **118**, 13801–13808 (2014).

23. Nguyen, L. *et al.* Chiral Assembly of Gold–Silver Core–Shell Plasmonic Nanorods on DNA Origami with Strong Optical Activity. *ACS Nano* **14**, 7454–7461 (2020).

24. Dass, M., Kuen, L., Posnjak, G., Burger, S. & Liedl, T. Visible wavelength spectral tuning of absorption and circular dichroism of DNA-assembled Au/Ag core–shell nanorod assemblies. *Mater. Adv.* **3**, 3438–3445 (2022).

25. Rycenga, M. *et al.* Controlling the Synthesis and Assembly of Silver Nanostructures for Plasmonic Applications. *Chem. Rev.* **111**, 3669–3712 (2011).

26. Ringe, E., Sharma, B., Henry, A.-I., Marks, L. D. & Duyne, R. P. V. Single nanoparticle plasmonics. *Phys. Chem. Chem. Phys.* **15**, 4110–4129 (2013).

27. Rührmair, U., Devadas, S. & Koushanfar, F. Security Based on Physical Unclonability and Disorder. in *Introduction to Hardware Security and Trust* (eds. Tehranipoor, M. & Wang, C.) 65–102 (Springer New York, NY, 2012).

28. Chen, H., Shao, L., Li, Q. & Wang, J. Gold nanorods and their plasmonic properties. *Chem. Soc. Rev.* **42**, 2679–2724 (2013).

29. Bohren, C. F. & Huffman, D.R. *Absorption and Scattering of Light by Small Particles* (John Wiley & Sons, Ltd, 1998).

30. Liu, J. J. *et al.* The Accurate Imaging of Collective Gold Nanorods with a Polarization-Dependent Dark-Field Light Scattering Microscope. *Anal. Chem.* **95**, 1169–1175 (2023).

31. Huang, Y. & Kim, D.-H. Dark-field microscopy studies of polarization-dependent plasmonic resonance of single gold nanorods: rainbow nanoparticles. *Nanoscale* **3**, 3228–3232 (2011).

32. Roller, E.-M. *et al.* Hotspot-mediated non-dissipative and ultrafast plasmon passage. *Nature Phys* **13**, 761–765 (2017).

33. Kershner, R. J. *et al.* Placement and orientation of individual DNA shapes on lithographically patterned surfaces. *Nature Nanotechnology* **4**, 557–561 (2009).

34. Martynenko, I. V. *et al.* Site-directed placement of three-dimensional DNA origami. *Nat. Nanotechnol.* 1–7 (2023).

## METHODS

### DNA Origami design, preparation and purification

The DNA origami nanodisc was designed using cadnano[35]. Design details of the nanodisc can be found in the supplementary information.

Staple strands were purchased from IDT Technologies (HPLC purified, 200 µM each in water). Cadnano files and list of sequences of oligonucleotides can be found in the Supplementary Data. The scaffold strands (p8634) were produced from M13 phage replication in Escherichia coli. DNA origami structures were folded by mixing scaffold strands with an excess of staple strands in folding buffer (buffers used in this work can be found in the 'DNA origami placement (DOP)' section of the supplementary information). Samples were annealed in a PCR machine (Tetrad 2 Peltier thermal cycler, Bio-Rad) and purified from excess staples by agarose gel electrophoresis. A full description of the folding and purification of the DNA origami can be found in the supplementary information.

**Preparation of DNA-coated AuNPs**

30 nm gold nanospheres were purchased from BBI International. All nanorods were prepared in-house. The nanoparticles were functionalized with a mixture of 5'-thiolated 19 nt and 8 nt poly-T single-stranded DNA (Biomers). The particles were centrifuged, the supernatant was removed and the particles were resuspended in 0.1% SDS to ~15 O.D. (for nanorods) and ~10 O.D. (nanospheres) concentration. The particles were then mixed with a 1:9 mixture of $T_{19}$ and $T_8$ thiolated single-stranded DNA (strand concentration 100 μM) in a nanoparticle suspension:DNA ratio of 3:2 (nanorods) or 3:1 (nanospheres) and frozen at -80° C for 30 min. After freezing, the functionalized particles were purified from excess DNA with gel electrophoresis in a 1% agarose gel ran for 90 min at 80 V. Particles were collected by cutting out the migrating band and squeezing out the liquid with the particles using a microscopy slide wrapped in parafilm. Specific staples in the nanodisc were modified with 10 nt poly-A extensions on the 3' ends (Fig. S1) to capture the nanoparticles.

**Preparation of the substrates and DNA origami placement**

Si/SiO2 substrates patterned with e-beam lithography were prepared by adapting the procedure from ref.[16] with slight modification. The 4-inch Si/SiO2 wafer with 100 nm thermal oxide (Microchemicals) was diced into 1 cm × 1 cm chips. Clean chips were primed with 10 mL of hexamethyldisilazane (HMDS) in a 4 L desiccator. The time of priming was optimised to maintain a Si/SiO$_2$ surface contact angle of 70°-75° after HMDS deposition. Binding sites were patterned into poly(methyl methacrylate) resist by electron-beam lithography. Examples of binding sites design are shown in figure S34. Then the chips were developed with a 1:3 solution of methyl isobutyl ketone (MIBK) and isopropanol (IPA). The HMDS in developed areas was removed with O2 plasma for 6 s in a plasma cleaner (PICO). The resist was stripped by ultrasonication in N-methyl pyrrolidone (NMP) at 50°C for 30 min. The substrates were briefly rinsed with 2-propanol, then dried in a nitrogen stream and used immediately.

Glass chips patterned via nanosphere lithography were prepared by adapting the procedure from ref.[17] 1 cm$^2$ glass chips were purchased from Plano-em. The polystyrene (PS) nanospheres with a diameter of 350 nm, 400 nm, 600 nm and 1μm (Thermo Scientific™ Nanosphere™ Size standards 3350A, 3400A, 3600A and 4010A respectively) were purified by centrifugation and resuspension in 50% ethanol/water. The PS nanospheres were drop-casted onto an O$_2$ plasma-activated chip surface and dried at a ~ 45° angle at RT, forming a close-packed monolayer/multilayer of nanospheres. The chips were then primed with 0.6 mL HMDS in a 1 L desiccator under a vacuum. The PS nanospheres were lifted off the surface by ultrasonication in water at RT for 5 min. Finally, the surface was blown dry with a nitrogen gun and baked at 120 °C for 5 min to stabilize the HMDS on the surface and used immediately.

Binding of DNA origami to the patterned substrates was achieved by depositing of a 60 μL drop of freshly folded and purified DNA origami in placement buffer to the surface of the chips. After incubation for 1 h at RT, excessive DNA origami was removed from the surface by 10 buffer replacement steps and purification with tween buffer. After this step, chips with placed DNA origami could be air dried, or nanoparticle placement could be performed. Air drying of the chips was performed by treating the samples with an ethanol dilution series (see supplementary information for step by step protocols protocols).

**Characterization techniques**

*UV-Vis spectroscopy.* Extinction measurements for determination of DNA and nanoparticle concentrations were performed with a NanoDrop ND-1000 spectrophotometer (Thermo Scientific).

*TEM imaging.* 5 µL of a sample was incubated for 30 s – 5 min, depending on concentration, on glow-discharged TEM grids (formvar/carbon, 300 mesh Cu; Ted Pella) at room temperature. DNA origami samples were stained with a 2% uranyl formate aqueous solution containing 25 mM sodium hydroxide. Imaging was performed with a JEM1011 transmission electron microscope (JEOL) operated at 80 kV.

*AFM imaging.* The tapping-mode AFM of glass substrates was carried out on a Dimension ICON AFM instrument (Bruker). OTESPA silicon tips (300 kHz, Vecco Probes) were used for imaging in air. Images were analysed with Gwyddion software.

*SEM imaging.* The SEM instrument used in this work is the Raith eLINE SEM instrument. The beam settings for imaging are 10 kV acceleration and 20 µm aperture. The samples were imaged using the SEM after 20 s sputtering using an Edwards Sputter Coater S150B. The sputter target contained 60% gold and 40% palladium. The process parameters used for sputtering were 5 mbar Argon, 1.5 kV, 11 mA. Here 20 s of sputtering results in the deposition of a layer of gold/palladium with a thickness of a few nanometres. SEM imaging was performed on horizontal samples.

**DATA AVAILABILITY**

Data available upon request from authors.

**METHODS-ONLY REFERENCES**

35. Douglas, S. M. *et al.* Rapid prototyping of 3D DNA-origami shapes with caDNAno. *Nucleic Acids Research* **37**, 5001–5006 (2009).

# Supplementary Information

## Methods

## DNA Origami

*Design.* A disc-shaped DNA origami structure was chosen to match and cover the circular binding sites created by the NSL process, specifically when using polystyrene nanospheres with $d \leq 400$nm. The cadnano design files and staple sequences are included separately. Staple extensions with 10 nt poly-A sequences were used to capture nanoparticles.

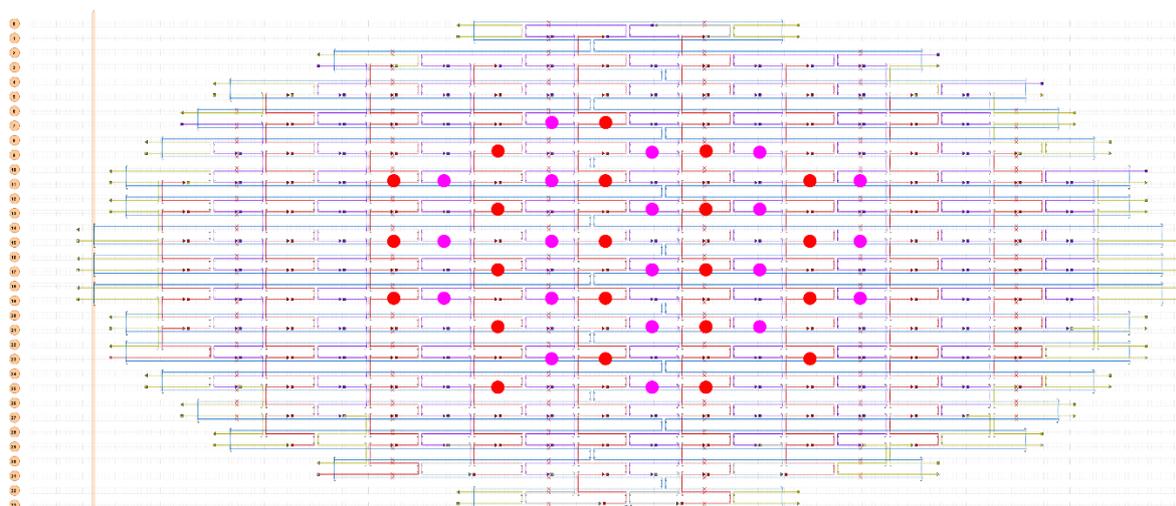

**Figure S1.** Cadnano design of the nanodisc origami structure. The scaffold is shown in blue. The red and violet staples are extendable from opposite faces. The green staples on the edges have $C_4$ extensions to prevent stacking. Deletions of base pairs are marked with red X symbols. Positions of staple extensions are marked by coloured circles.

*Assembly.* Staple strands (Integrated DNA Technologies, 200 µM each in water) and the scaffold strand (modified 8634 nt long M13mp18 ssDNA) were mixed to a target concentration of 200 nM for each staple and 20 nM for the scaffold, respectively, in 18 mM $MgCl_2$, 10 mM Tris and 1 mM EDTA (pH 8.3). The mixture was divided into 100 µL aliquots in PCR tubes and annealed from 65ºC to 20ºC over ~ 16 hours (see table below).

***Table S1.*** **DNA Origami folding program**

| Temperature (ºC) | Time |
|---|---|
| 65 | 15 min |
| 64-60 | 5 min/ºC |
| 59-40 | 45 min/ºC |
| 39-36 | 30 min/ºC |
| 35-20 | 5 min/ºC |

*Purification.* The origami structures were analysed and purified using a 1% agarose gel. The gels were run in a buffer containing 40 mM Tris, 20 mM acetic acid, 1 mM EDTA and 11 mM MgCl2 at 100V for 2 hours. 1x SybrSafe (Thermo Fisher) was included in gels for visualizing DNA. The desired bands were excised and squeezed between parafilm sheets to extract the purified sample. We do not believe the SybrSafe affects further downstream experiments.

### *Characterisation*

*UV-Vis spectroscopy.* Extinction measurements for determination of DNA and nanoparticle concentrations were performed with a NanoDrop ND-1000 spectrophotometer (Thermo Scientific).

*TEM imaging.* 5 µL of a sample was incubated for 30 s – 5 min, depending on concentration, on glow-discharged TEM grids (formvar/carbon, 300 mesh Cu; Ted Pella) at room temperature. DNA origami samples were stained with a 2% uranyl formate aqueous solution containing 25 mM sodium hydroxide. Imaging was performed with a JEM1011 transmission electron microscope (JEOL) operated at 80 kV.

*AFM imaging.* The tapping-mode AFM of glass substrates was carried out on a Dimension ICON AFM instrument (Bruker). OTESPA silicon tips (300 kHz, Vecco Probes) were used for imaging in air. Images were analysed with Gwyddion software.

*SEM imaging.* The SEM instrument used in this work is the Raith eLINE SEM instrument. The beam settings for imaging are 10 kV acceleration and 20 µm aperture. The samples were imaged using the SEM after 20 s sputtering using an Edwards Sputter Coater S150B. The sputter target contained 60% gold and 40% palladium. The process parameters used for sputtering were 5 mbar Argon, 1.5 kV, 11 mA. Here 20 s of sputtering results in the deposition of a layer of gold/palladium with a thickness of a few nanometers. SEM imaging was performed on horizontal samples.

## Plasmonic Nanoparticles

*Synthesis.* The gold nanorods were synthesised as in ref.[1]. The procedure was optimised until the desired size was achieved. The rods were washed in 0.1 M CTAB (Roth) and 0.01 M CTAB and stored in 0.01 M CTAB at 15 O.D. (optical density) before use. The gold nanospheres (BBI solutions) were centrifuged and washed into 0.1% SDS @ 10 O.D. before use.

*Silver coating.* The silver coating was performed as previously reported.[2] The nanoparticles were added to 0.1 M CTAB under stirring (500 rpm) and allowed to mix for 10 s. Then, a 1:9 mixture of thiolated ssDNA strands with a 19 nt long poly-T sequence (T19) and an 8 nt long poly-T sequence (T8) (Biomers, 100 µM, aq.), respectively, were added to the rod mixture. This was followed by adding $AgNO_3$ solution, after which the stirring speed was increased (1500 rpm). L-Ascorbic acid (0.2 M) and NaOH (0.2 M) were added rapidly in quick succession. A colour change after 5 - 10 s indicates the successful synthesis of a silver coating. The reaction was allowed to proceed for 10 min. The core-shell nanoparticles were washed to remove excess reactants from their respective growth solutions and then redispersed in 0.1% Sodium dodecyl sulfate (SDS) before further use.

*Table S2.* Silver-coating protocol

| Reagent | Concentration | Volume |
|---|---|---|
| CTAB | 0.1 M | 2250 uL |
| AuNR<br><br>**AuNS** | 15 O.D. in 0.01 M CTAB<br><br>**6 O.D. in 0.01 M CTAB** | 480 uL |
| Thiol-DNA | T8:T19 (9:1) 100 uM in water | 480 uL |
| AgNO$_3$ | 8 mM  **(6 mM for AuNS)** | 400 uL |
| L-Ascorbic acid | 0.2 M | 64 uL |
| NaOH | 0.2 M | 125 uL |

*Washing AuAgNRs*

1. The synthesised rod dispersion was aliquoted into 300 uL batches.
2. The rods were centrifuged and then redispersed in 0.1M CTAB. Centrifugation parameters were 4500 rcf, 5 min for rods synthesised with 2-6 mM AgNO$_3$, and 3000 rcf, 3 min for rods synthesised with 8-12 mM AgNO$_3$.
3. The rods were centrifuged and then redispersed in 0.01 M CTAB.
4. The rods were centrifuged and then redispersed in 0.1% SDS.

**Note:** Aggregation of the AuAgNRs most commonly occurred during the washing steps. If the centrifugation speed was too high, the AuAgNR pellet had a metallic lustre and was challenging to redisperse. For washing, we advise starting from lower centrifugal speeds and gradually increasing them until a clear supernatant is obtained.

*Functionalisation.* Nanorods were resuspended in 0.1% SDS to ~15 O.D. before functionalisation with DNA. In a typical experiment, 32 uL of thiol-T19 and 288 uL of thiol-T8 (1:9 ratio) were mixed and added to 480 uL nanorods dispersed in 0.1% SDS. Nanospheres were resuspended in 0.1% SDS to ~10 O.D. before functionalisation with DNA. In a typical experiment, 16 uL of thiol-T19 and 144 uL of thiol-T8 were mixed and added to 480uL nanospheres dispersed in 0.1% SDS.

In all cases, the mixture was vortexed for 5 seconds and kept at -80ºC for 30 min. The mixture was then thawed and centrifuged to concentrate it to ~100 uL. Agarose gel electrophoresis was used to separate the functionalised nanoparticles from excess thiol-DNA.

*Purification.* The nanoparticles were analysed and purified using a 1% agarose gel. The gels were run in a buffer containing 40 mM Tris, 20 mM acetic acid, 1 mM EDTA and 11 mM MgCl$_2$ at 100V for 90 minutes. The desired bands were excised and squeezed between parafilm sheets to extract the purified sample.

*Characterisation.*

*UV-Vis spectroscopy*. 1.5 uL of the liquid on the substrate was taken to measure the nanoparticle concentration during NPP.

*TEM imaging*. See the DNA origami section. No staining was performed.

# DNA origami placement (DOP)

The DNA origami placement procedure was adapted from ref.[3].

*Buffers used*

*Folding buffer.* 10 mM Tris, pH 8.3, 1 mM EDTA, 18 mM $MgCl_2$.

*Placement buffer.* 40mM Tris-HCl, pH 8.35, 40mM $MgCl_2$.

*Tween buffer.* 40mM Tris-HCl, pH 8.35, 40mM $MgCl_2$, 0.07% Tween20.

*Gel buffer.* 40mM Tris base, 20mM acetic acid, 1mM EDTA, pH 8.3, 11mM $MgCl_2$.

*Substrate preparation and binding site fabrication.*

*Materials and equipment required:-*

- Polystyrene nanospheres (diameter - 350nm, 400nm, 600nm, 1μm) at 1 wt% concentration (Distrilab)
- 10mm x 10mm coverslips (Plano-em)
- Lobind tubes
- Tweezers (Plano-em)
- Hotplate
- Plasma cleaner (Diener Pico)
- Ultrasonication bath
- Isopropanol (Roth)
- HMDS
- Custom 3D-printed stands to hold the coverslips at a 45° angle.
- Dessicator (1L)

1. Wash ~ 500uL polystyrene nanospheres (PsNs) by centrifuging, removing the supernatant and resuspending them in 50% Ethanol/water (v/v) three times.
2. Resuspend PsNs to 3.5 wt% (~ 150 uL) in 50% Ethanol/water (v/v).
3. Mark a small scratch on the coverslips to ensure upright orientation in the event of them flipping over during the binding site fabrication process.
4. Sonicate coverslips in isopropanol for two minutes @ 100% power.
5. Place coverslips in a petri dish and plasma clean @ 50W power, 5 min, 45sccm $O_2$ flow.
6. Place the coverslips against the 3D-printed stand to maintain a reproducible 45° angle.

7. Drop 8uL of the polystyrene nanosphere suspension to a coverslip. The plasma treatment should render the coverslips hydrophilic, and the suspension should spread over the whole surface.
8. Wait until the suspension dries completely, around 5 minutes.
9. Heat at 60°C for 5 minutes to dry the coverslips.
10. Place the coverslips in a petri dish and plasma clean @ 20W power, 2 min, 45sccm $O_2$ flow. This is to clean the surface again (so-called 'descum' step) and render the surface hydrophilic again.
11. Add 600uL of HMDS (in a small cup) in the desiccator. Place the coverslips on a plate above the cup, and deposit under a vacuum seal (40mbar) for 20 minutes.
12. Lift-off PsNs by immersing coverslips in DI water and ultrasonicating @100% power for 5 minutes.
13. Heat at 120°C for 5 minutes to stabilise the HMDS-treated surface.

### *PUF fabrication.*

*Materials and equipment required:-*

- Patterned chips
- Placement buffer
- Tween buffer
- Purified DNA Origami
- Purified DNA-functionalized nanoparticles
- Petri dish
- Parafilm
- Tweezers
- Pipette and tips
- Kimtech wipes
- Ethanol
- DI water

*DNA Origami Placement*

1. Place a small strip of parafilm inside a Petri dish.
2. Place the patterned chip on the parafilm. The hydrophobic parafilm often pushes the buffer droplet back onto the chip in the event of spillage.
3. Keep a moistened Kimtech wipe in the Petri dish to help maintain humidity during incubation.
4. Dilute the purified DNA Origami with *Placement buffer* in LoBind tubes.
5. Pipette ~ 60 uL of the diluted Origami onto the chip. **Incubate for 1 hour**.
6. Wash 10x with the *Placement buffer* by pipetting 60uL fresh buffer onto the chip, mixing 2-3 times and pipetting 60 uL off the chip.

7.  Wash 5x with the *Tween buffer* by pipetting 50 uL fresh buffer onto the chip, mixing 2-3 times and pipetting 50 uL off the chip. Addition of the *Tween buffer* results in a change of the surface tension and spreading of the droplet. **Incubate for 5 minutes**.
8.  Wash away the *Tween buffer* by washing for 5 minutes (~ 30 washes) with the *Placement buffer* by pipetting 80 uL fresh buffer onto the chip, mixing 2-3 times and pipetting 80 uL off the chip. The indication for the Tween being washed away is that the droplet returns to its original shape.

If the experiment aims to only place DNA Origami, for example, for imaging or troubleshooting purposes, skip to step 15 for the drying procedure.

*Adding nanoparticles*

9.  Centrifuge the Purified DNA-functionalized nanoparticles. Remove as much supernatant as possible, and then vortex and sonicate the suspension to redistribute the nanoparticles uniformly.
10. Now take the desired amount of this nanoparticle suspension (usually 1-10 uL) and dilute it with the *Placement buffer*.
11. Take away 30 uL of the *Placement buffer* on the chip. Add 60 uL of the nanoparticle suspension in *Placement buffer* to the chip, mixing 2-3 times and leaving the liquid on the chip. **Incubate for 1 hour.**
12. Take away 30 uL of the nanoparticle suspension on the chip. Wash 10x with the *Placement buffer* by pipetting 60 uL fresh buffer onto the chip, mixing 2-3 times and pipetting 60 uL off the chip.
13. Wash 5x with the *Tween buffer* by pipetting 50 uL fresh buffer onto the chip, mixing 2-3 times and pipetting 50 uL off the chip. Addition of the *Tween buffer* results in a change of the surface tension and spreading of the droplet. **Incubate for 5 minutes**.
14. Wash away the *Tween buffer* by washing for 5 minutes (~ 30 washes) with the *Placement buffer* by pipetting 80 uL fresh buffer onto the chip, mixing 2-3 times and pipetting 80 uL off the chip.

*Drying procedure*

15. Successively place the chip into 25%, 50%, 75% and 85% Ethanol/water (v/v) mixture for 10 seconds, 10 seconds, 20 seconds and 2 minutes, respectively.
16. Air dry the chips on a Kimtech wipe.

**Note:** All the DNA Origami and nanoparticles in this work were gel purified for the experiments. The rationale behind that is twofold - to get rid of any spurious assemblies (dimers, aggregates) and to ensure the removal of all excess ssDNA strands, both staples and thiol-DNA. The latter is essential because free thiol-DNA could bind to anchors and inactivate DNA Origami structures placed on the surface. Other purification procedures might result in similar assembly quality.

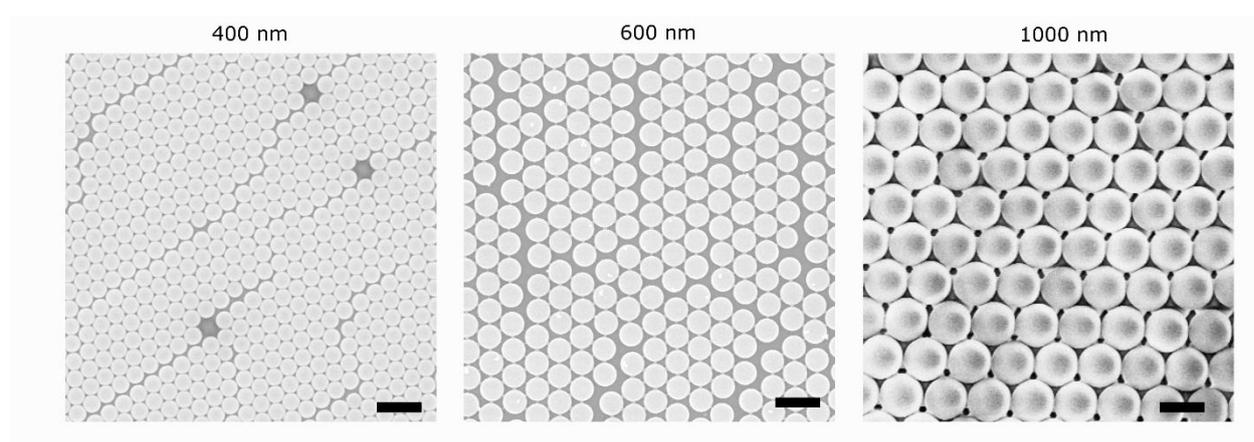

**Figure S2.** SEM images of polystyrene nanospheres deposited on the glass surface. Numerous defects (point, line) are clearly visible. Also noteworthy is that since the nanospheres are deposited using tilted drop-casting, we obtain multilayers of the nanospheres along with monolayers, which has no adverse effect on the DOP. Scale bars: 1 μm.

***Table S3.*** **Theoretical comparison of the surface area of a single nanodisc relative to the binding sites created using NSL.**

Results were calculated using $diameter_{BS}=0.27x$ $diameter_{nanosphere}$ [4].

|  | Surface Area ($nm^2$) | $N_{DNO}$ per BS (theoretical) |
|---|---|---|
| DNO (Nanodisc) | ~7000 | - |
| Binding site with 350nm nanospheres ($d_{BS}$=95nm) | ~7000 | ~1 |
| Binding site with 400nm nanospheres ($d_{BS}$=110nm) | ~9500 | 1-2 |
| Binding site with 600nm nanospheres ($d_{BS}$=160nm) | ~20000 | 3-4 |
| Binding site with 1000nm nanospheres ($d_{BS}$=270nm) | ~57000 | 8-9 |

## Supplementary Note S1: Nanoparticles

Four types of nanoparticles were used: 50 nm Silver-coated gold nanospheres (AgNS), 30 nm gold nanospheres (AuNS), silver-coated gold nanorods (AgNR) with length = 75 nm and diameter = 25 nm and gold nanorods (AuNR) with length = 60 nm and width = 12 nm.[2,5]

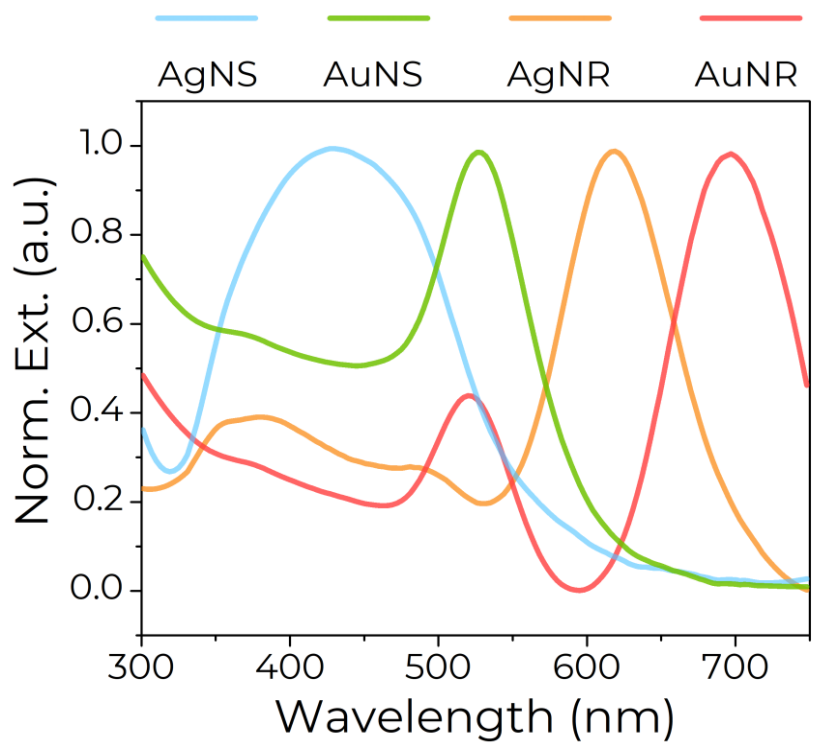

**Figure S3.** Normalized extinction spectra of the nanoparticles used, acquired using UV-vis spectroscopy.

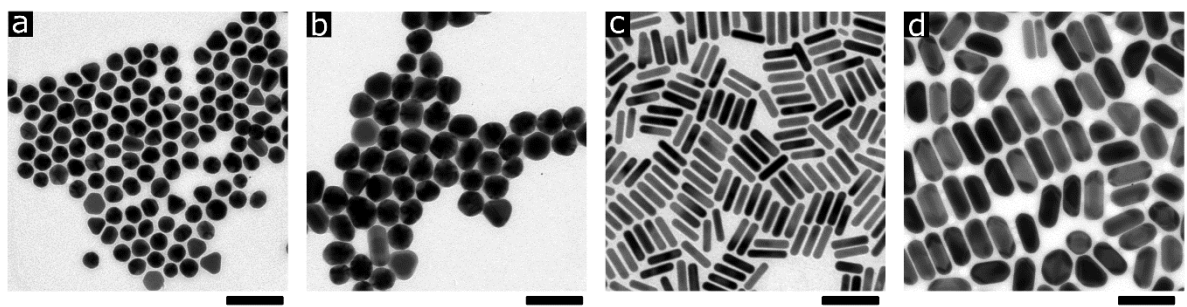

**Figure S4.** TEM images of a) AuNS, b) AgNS, c) AuNR and d) AgNR. Scale bars : 100 nm. The particles were dried from water suspension on glow-discharged TEM grids (formvar/carbon, 300 mesh Cu; Ted Pella).

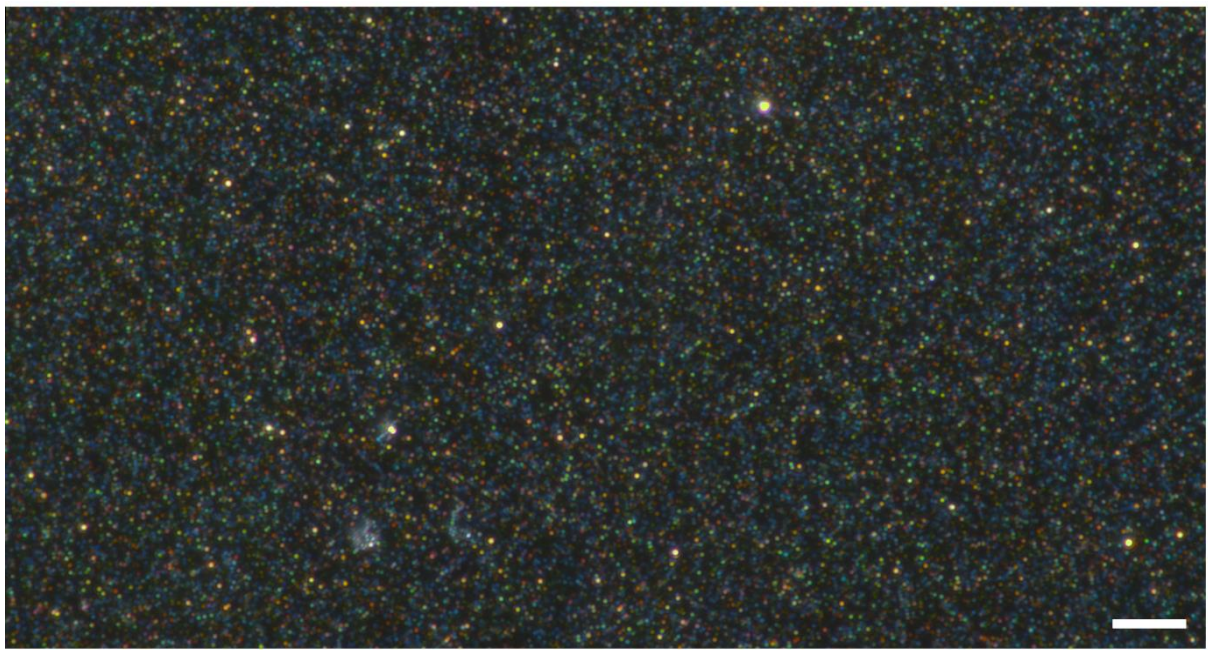

**Figure S5.** Uncropped versions of the PUFs in Fig. 2a (400 nm spacing) recorded at 100x magnification. Scale bar, 10 μm.

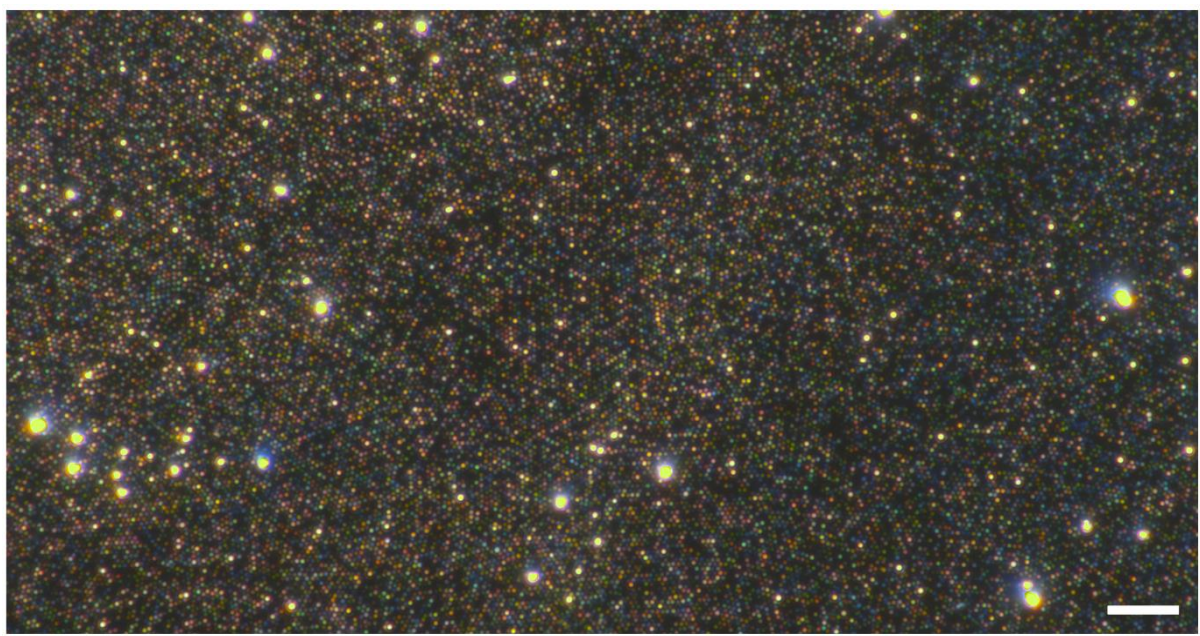

**Figure S6.** Uncropped versions of the PUFs in Fig. 2b (600 nm spacing) recorded at 100x magnification. Scale bar, 10 μm.

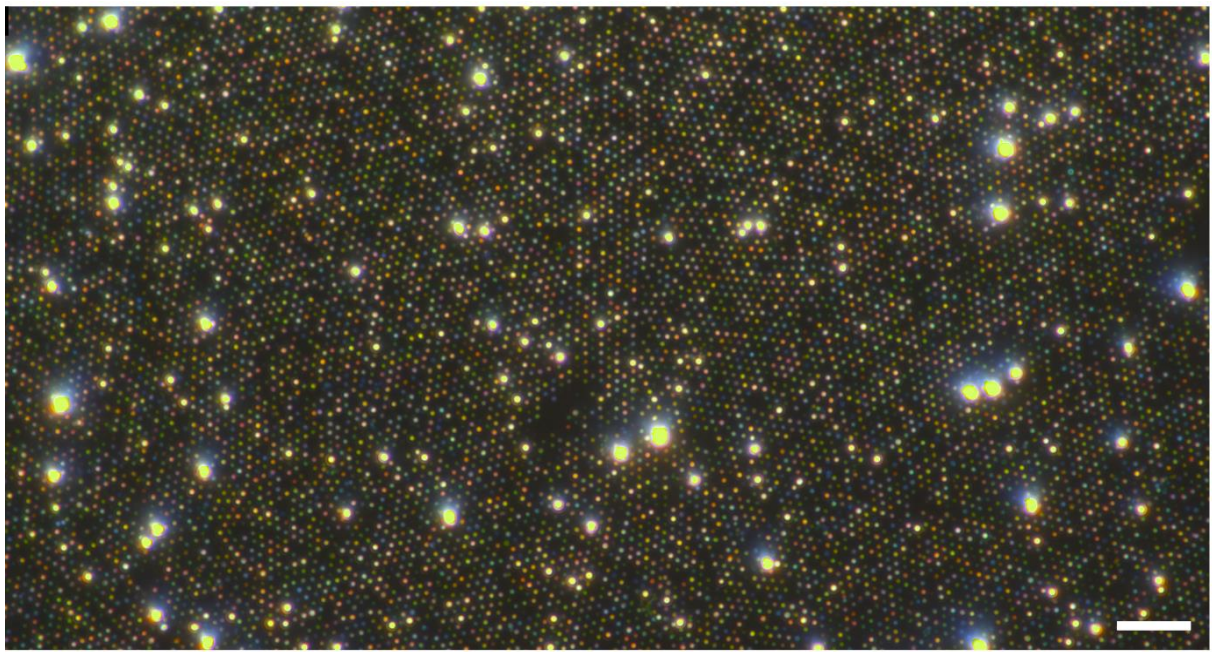

**Figure S7.** Uncropped versions of the PUFs in Fig. 2c (1 um spacing) recorded at 100x magnification. Scale bar, 10 μm.

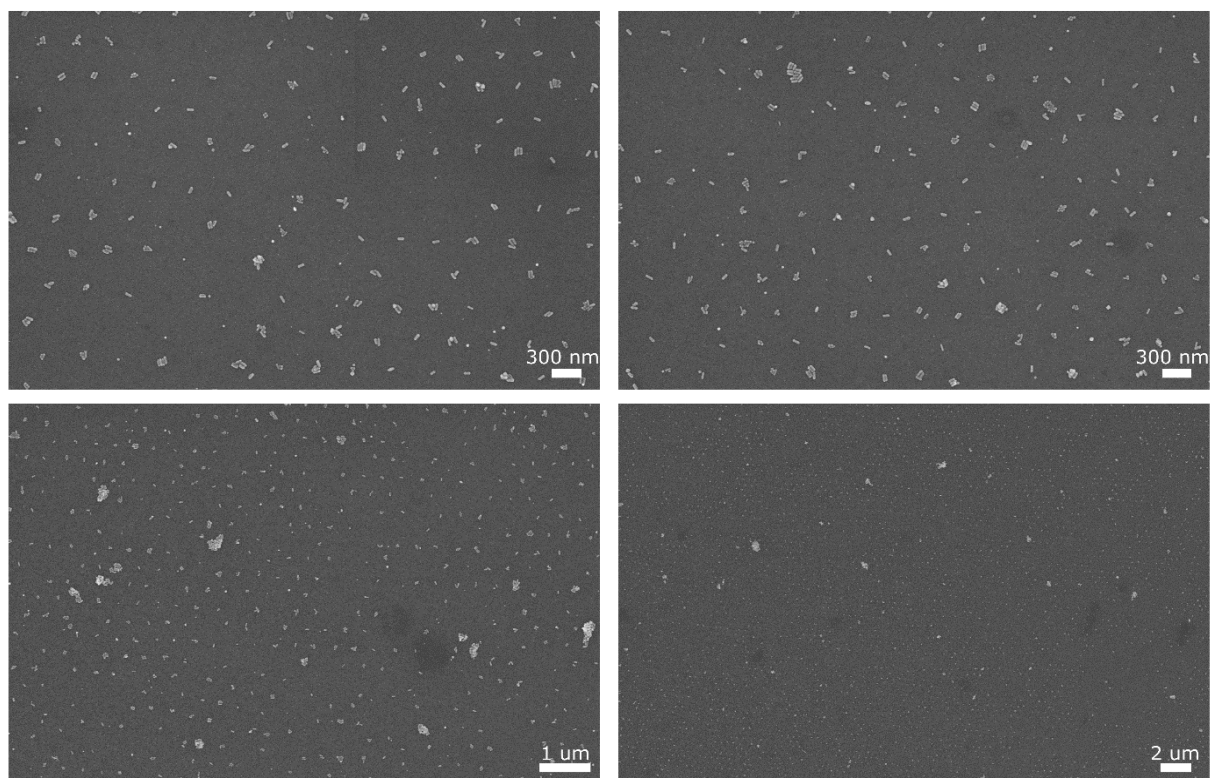

**Figure S8.** SEM images of a PUF with 400 nm spacing. Particles used – AuNS, AuNR, AgNS, AgNR @ 0.5 OD.

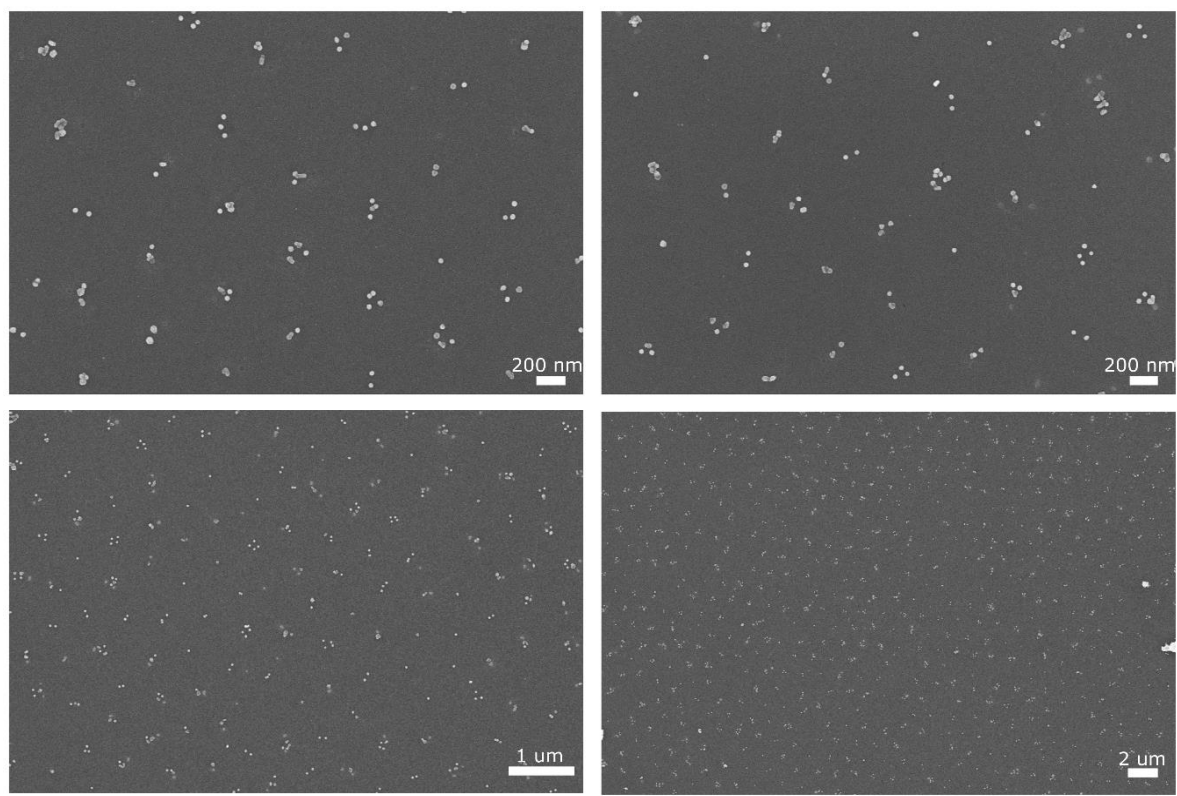

**Figure S9.** SEM images of a PUF with 600 nm spacing. Particles used – AuNS, AgNS, AgNR @ 0.2 OD.

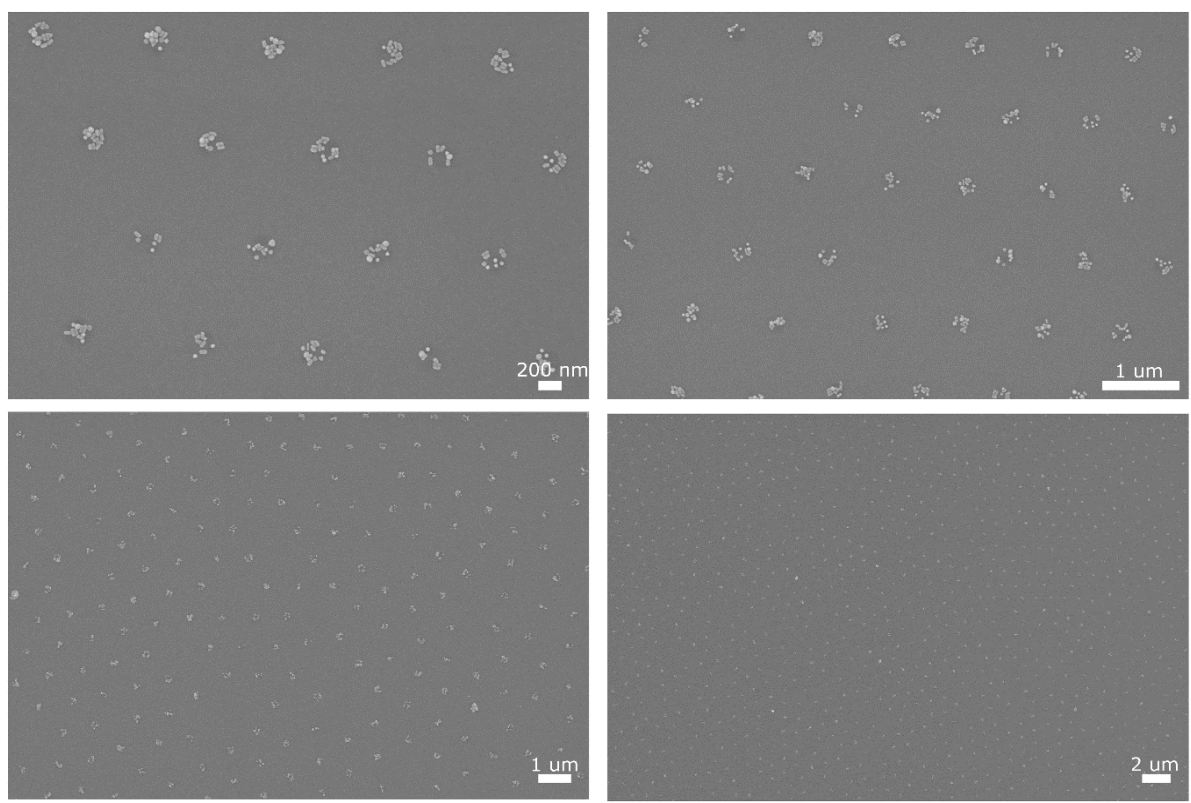

**Figure S10.** SEM images of a PUF with 1 um spacing. Particles used – AuNS, AgNS, AgNR @ 0.25 OD.

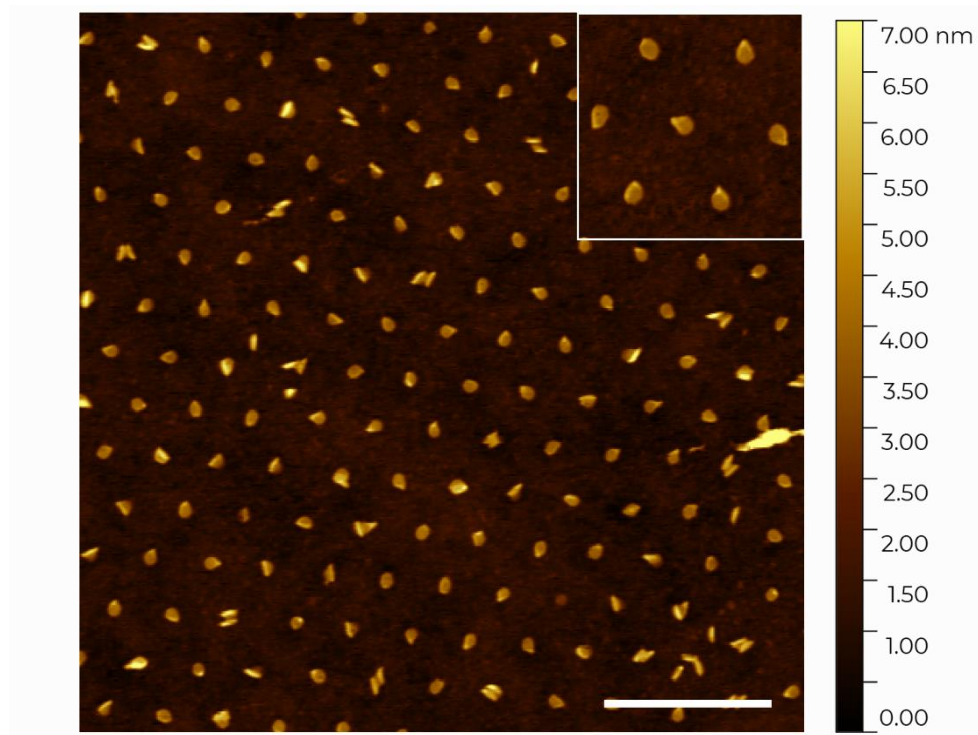

**Figure S11.** Optimised DOP. NSL was performed using 350 nm polystyrene nanospheres. DNA origami concentration during DOP was 150 pM. Scale bar, 1 μm. Inset: 800 nm x 800 nm.

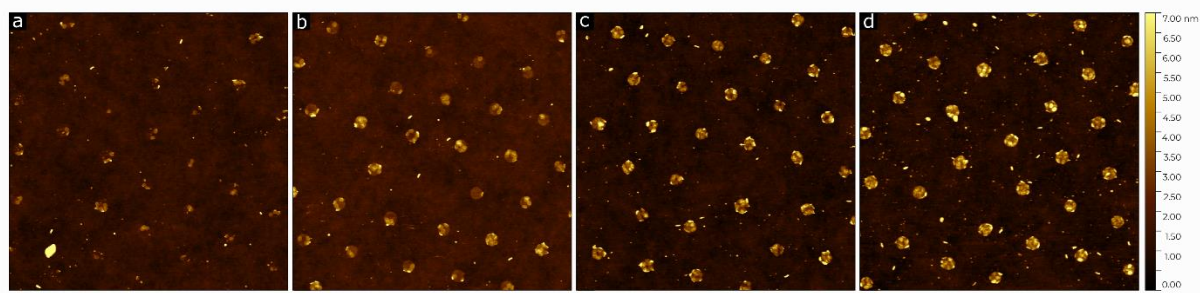

**Figure S12.** NSL was performed using 1um nanospheres. DNA origami concentration during DOP was - a. 50 pM, b. 100 pM, c. 225 pM and d. 450 pM. All scans are 6x6 μm.

## Supplementary Note S2: Image analysis pipeline

1. Input: RGB .tif file (here: 4096x2160px)
2. Image processing (Fig. S13)
3. Particle measurement (x, y, hue, saturation, value)
4. Output: .csv file with particle attributes
5. Plot hue histograms

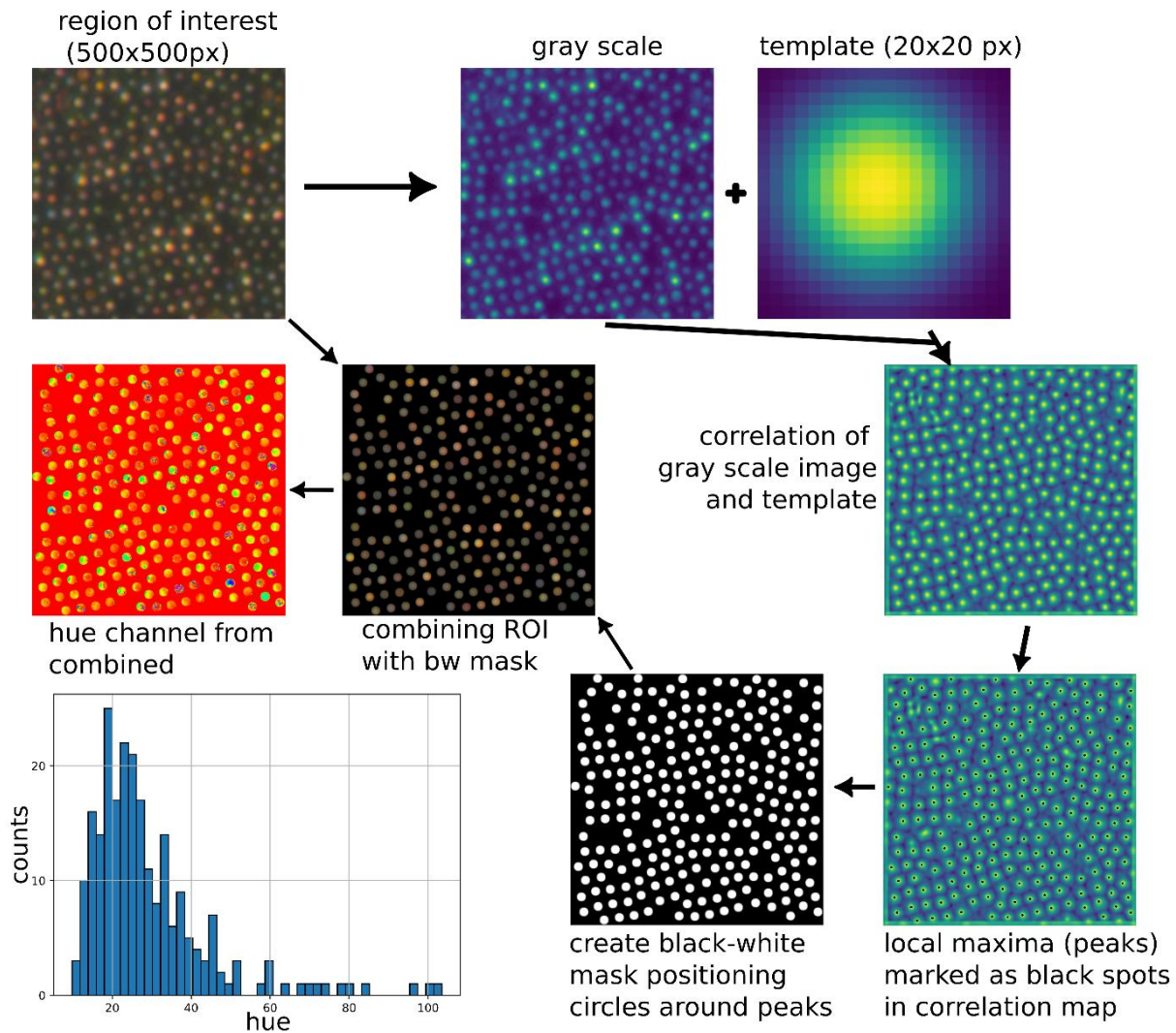

**Figure S13.** Image processing workflow. First, transform the initial RGB image to grayscale. Use a dummy particle as template to get a correlation map. Get local maxima for areas where correlation is higher than a certain value (usually 0.8- 0.9, can also be 0.5 for 600 nm patterning). Draw a circle around the coordinates of the peaks to get the black-white mask. Combine the mask with the RGB image to get visual feedback of where particle data is harvested. To get the hue histogram, the RGB image has to be converted to HSV colorspace and then split into 'hue', 'saturation' and 'value' channels. The hue value of one particle is the mean hue of all pixels in this particle.

Data pipeline for obtaining hue from polarised images :

1. Input: .avi file.
2. Crop to the desired size.
3. Image processing of each frame.
4. Particle measurement and rearrangement to polarisation data for each particle.
5. Plot particle hue/saturation/value – polarisation for individual particles (Fig. S14).

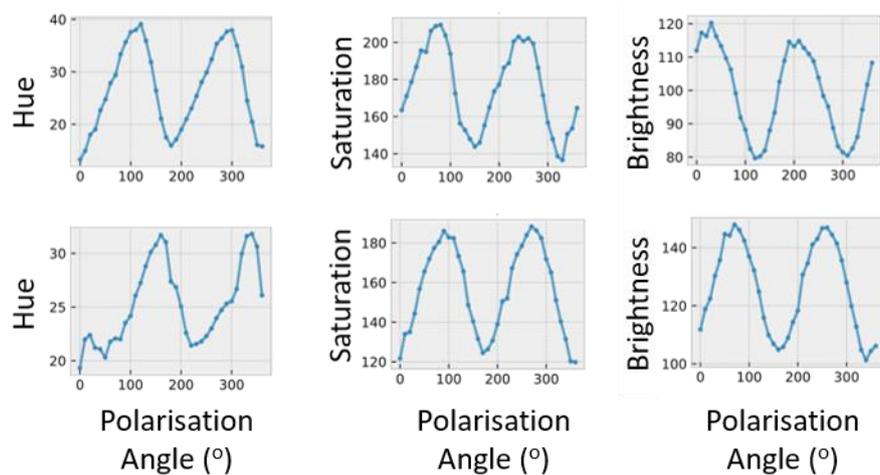

**Figure S14.** Plots showing the variation of hue, saturation and brightness versus polarisation for two particles provided as output by the image analysis algorithm.

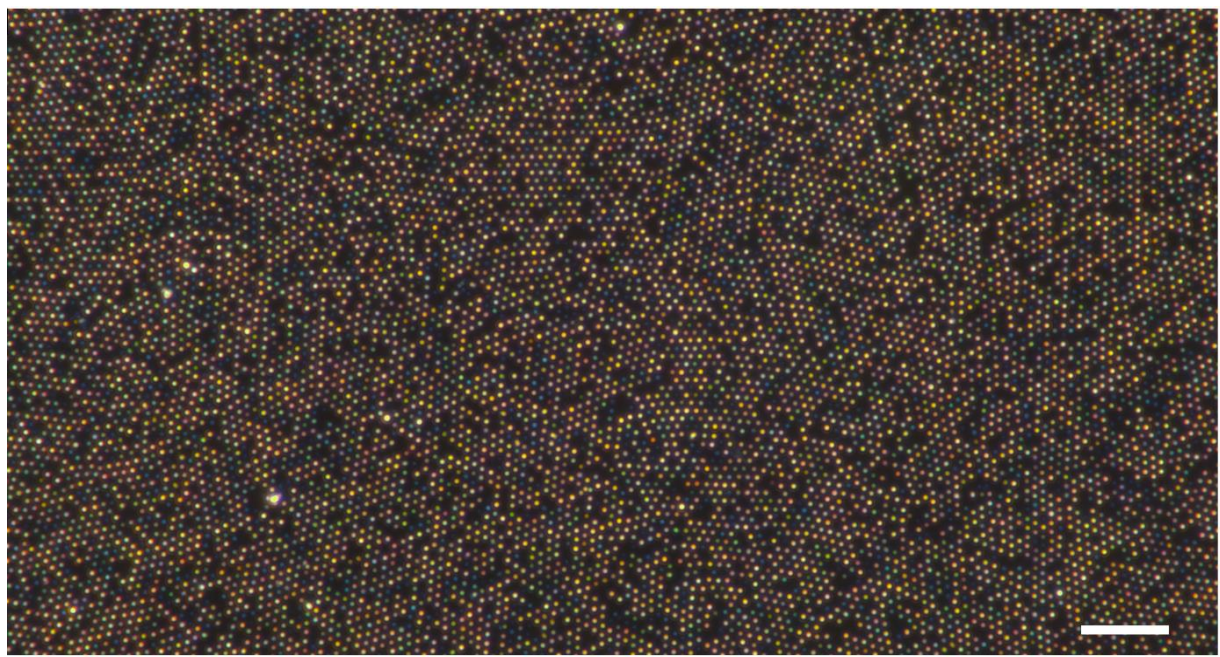

**Figure S15.** Uncropped versions of the PUFs in Fig. 3a: AgNS (high concentration) recorded at 100x magnification. Scale bar, 10 µm. The spacing is 1 µm.

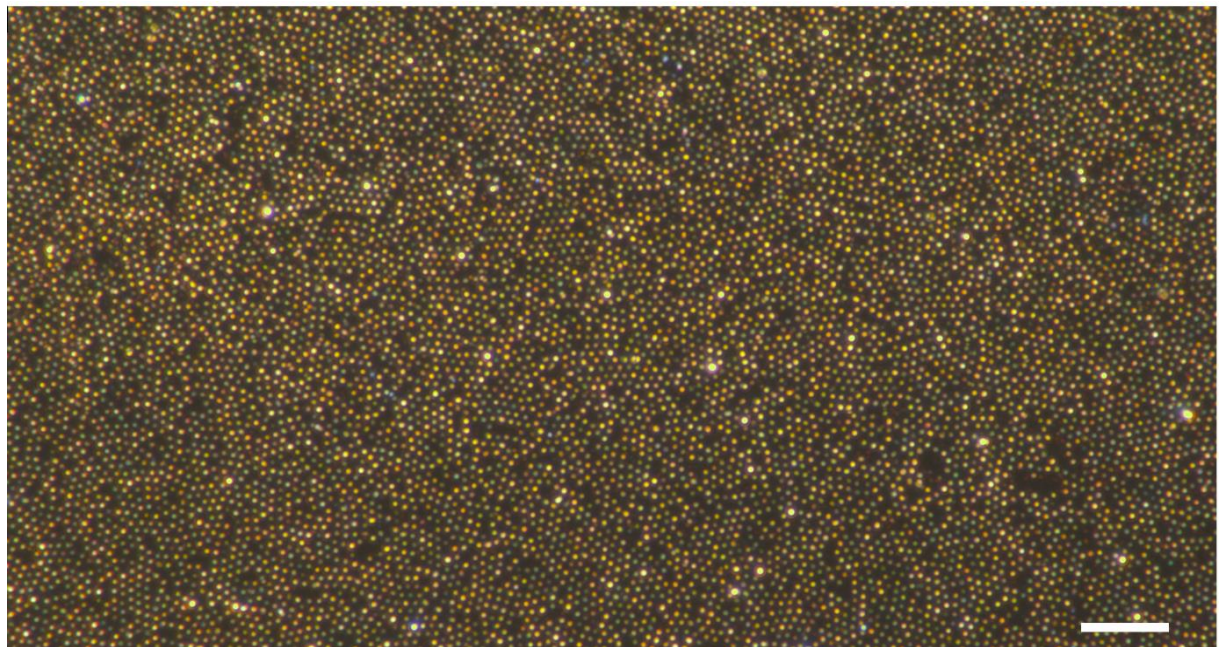

**Figure S16.** Uncropped versions of the PUFs in Fig. 3b: AuNS+AuNR (high concentration) recorded at 100x magnification. Scale bar, 10 μm. The spacing is 1 μm.

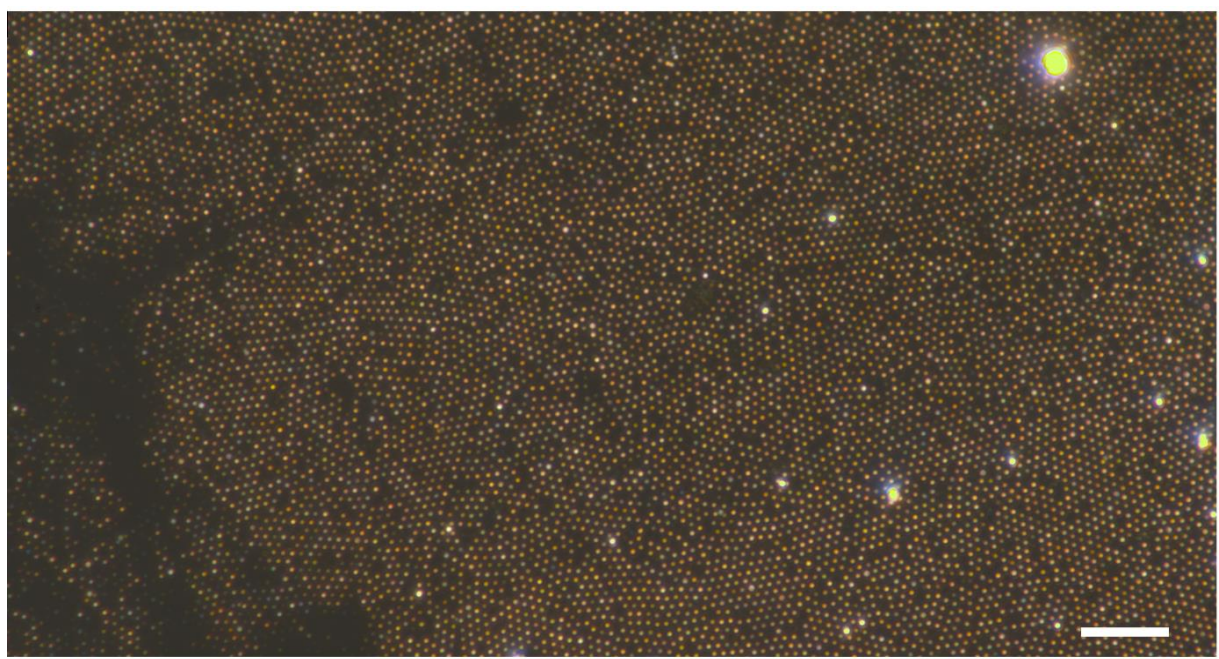

**Figure S17.** Uncropped versions of the PUFs in Fig. 3c: AuNS+AgNS+AgNR (high concentration) recorded at 100x magnification. Scale bar, 10 μm. The spacing is 1 μm. The area in the bottom left was intentionally scratched with a pipette tip to locate the region during SEM imaging.

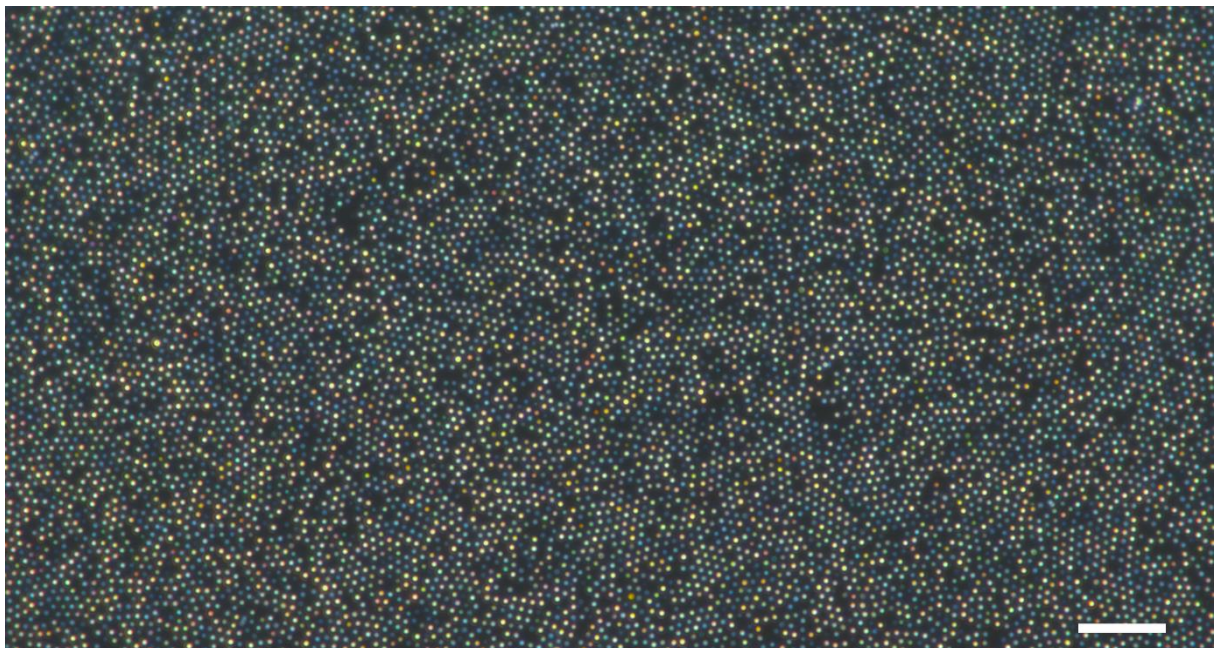

**Figure S18.** Uncropped versions of the PUFs in Fig. 3d: AgNS (low concentration) recorded at 100x magnification. Scale bar, 10 µm. The spacing is 1 µm.

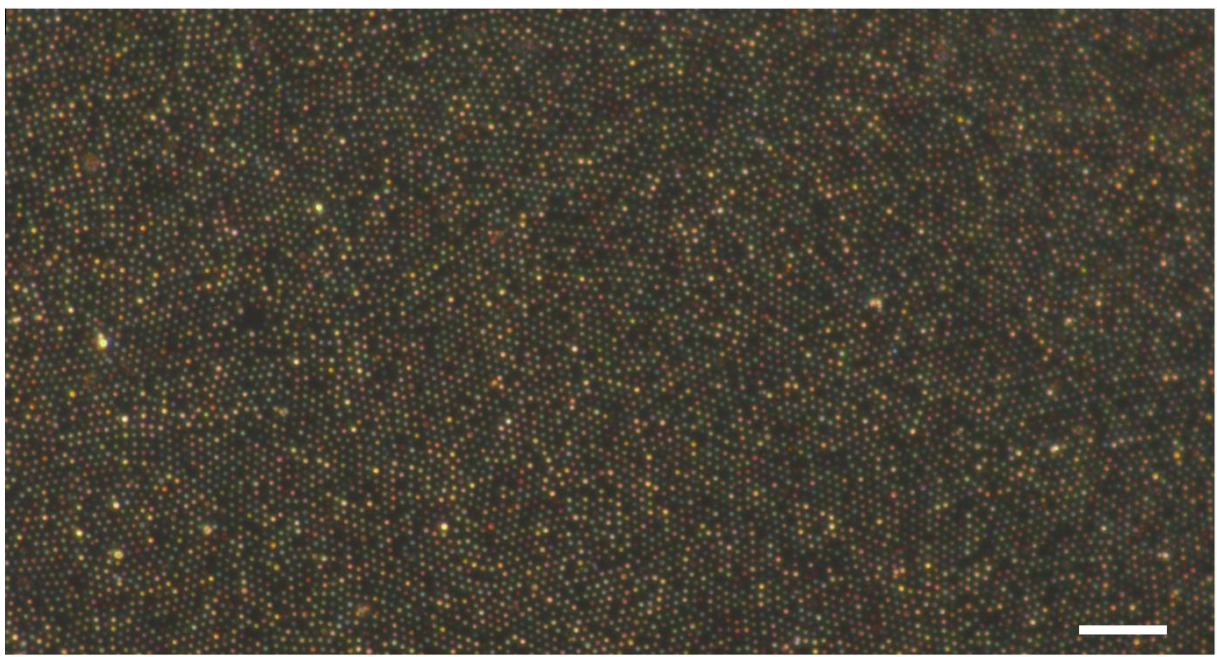

**Figure S19.** Uncropped versions of the PUFs in Fig. 3e: AuNS+AuNR (low concentration) recorded at 100x magnification. Scale bar, 10 µm. The spacing is 1 µm.

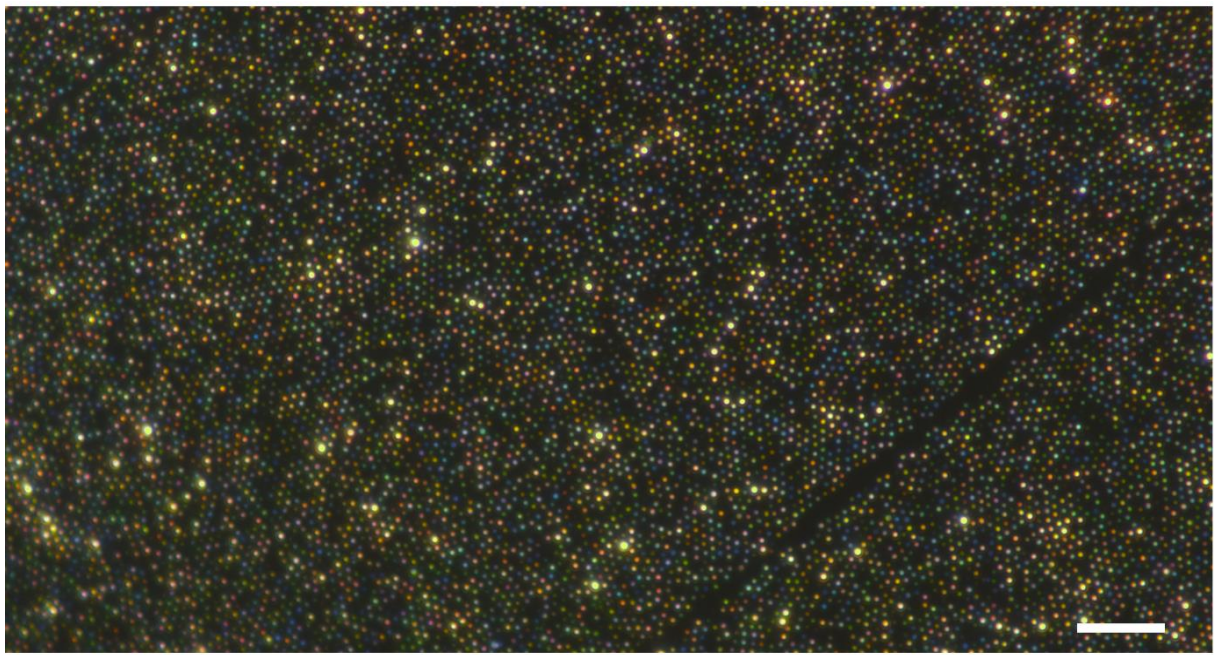

**Figure S20.** Uncropped versions of the PUFs in Fig. 3f: AuNS+AgNS+AgNR (low concentration) recorded at 100x magnification. Scale bar, 10 µm. The spacing is 1 µm.

*Table S4. Placement conditions for Fig. 3*

| Sample | Particle mix | DNO conc. (pM) | Particle conc. (O.D.) |
|--------|--------------|----------------|------------------------|
| a | 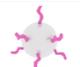 | 40 | 0.75 |
| b | 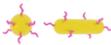 | 40 | 1.5 |
| c | 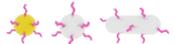 | 25 | 2.7 |
| d | 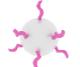 | 40 | 0.5 |
| e | 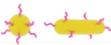 | 40 | 0.5 |
| f | 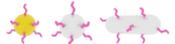 | 40 | 0.25 |

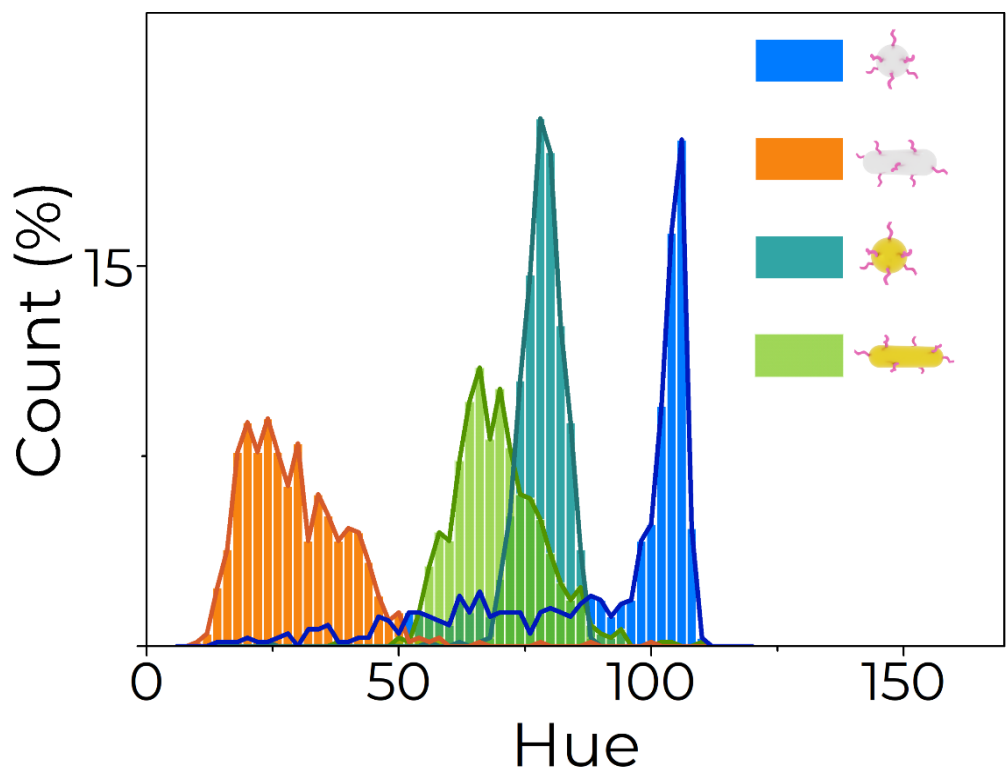

**Figure S21.** Hue histograms of the nanoparticles used in this study. Each variety of nanoparticle was suspended in water, deposited on separate plasma cleaned glass slides at O.D. ~ 0.1 and incubated for 30 s. The slides were then dried with $N_2$ flow and imaged using DFM (Fig. S28-31). Nanorods have a tendency to stack along their long-axis, forming a variety of aggregates.[6] This results in a wider distribution of hues. Multi-particle assemblies scatter at longer wavelengths and therefore generally increased the frequency of hues of lower values on the HSV scale. Therefore, it might be possible to increase the range of achievable hues by using particles that scatter at higher hue values.

## Supplementary Note S3: Variation of extensions

DNA origami provides the unique opportunity to control the number of extensions protruding from the structures. To verify the effect of this influence, we performed DOP using nanodiscs with varying number of extensions. Only gold and silver nanospheres were used in these PUFs (O.D.~0.8). We varied the number of extensions from 44 (22 from each face), to 8 (4 from each face) to 0. Using a larger number of extensions resulted in a broad hue response, consistent with stochastic assembly of nanoparticles at each placement site (Fig. S22a). A lower number of extensions gave a narrower hue distribution, attributed to the capture of mainly single particles along with small aggregates (Fig. S22b). As expected, using DNA Origami structures without extensions resulted in surfaces with a much lower degree of NPP, with no apparent lattice visible under Dark Field illumination (Fig. S22c). The number of extensions thus directly affects the hue response of the PUFs.

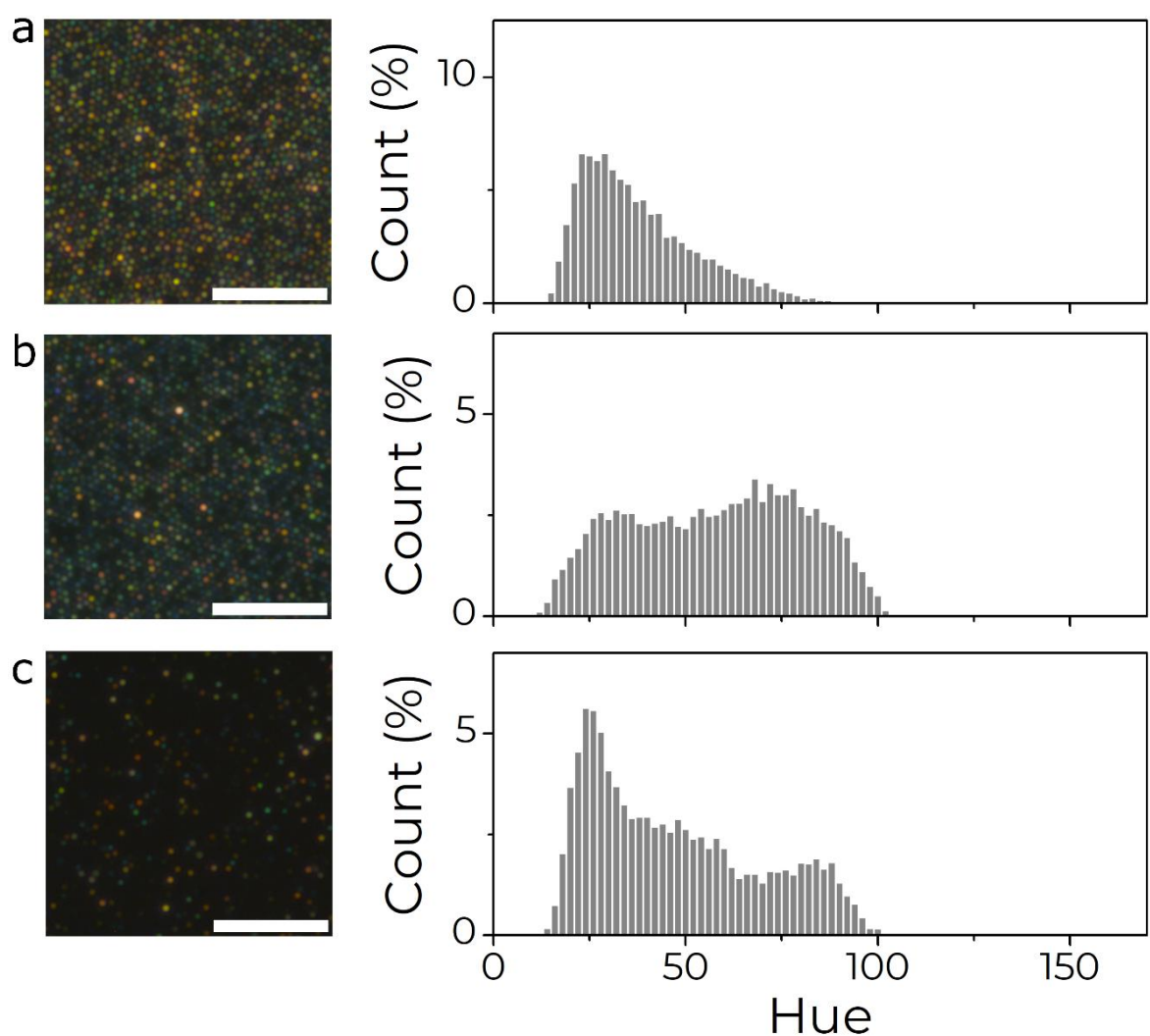

**Figure S22.** The number of extensions from the DNO placed are a) 44 ($d_{binding\ area}$=60 nm), b) 8 ($d_{binding\ area}$=10 nm) and c) 0. Placement parameters: Polystyrene nanospheres-600 nm, $C_{DNO}$ = 40pM, $C_{NP}$= 0.8 O.D. Scale bars, 10 μm. N=10000 spots analysed for all histograms.

*Table S5. Dependence of hue on imaging conditions*

The DFM uses an Olympus Model U-LH100-3 lamp @12V, 100W. The table below shows the variation in hue of a single spot under a variety of imaging conditions.

| Hue | Δ % | Illumination level | Exposure times |
|---|---|---|---|
| 20 (standard imaging conditions) | - | 100% | 1000ms |
| 18 | 10 % | 100% | 500ms |
| 19 | 5 % | 100% | 700ms |
| 22 | 10 % | 100% | 900ms |
| 21 | 5 % | 66% | 400ms |
| 20 | 0 % | 83% | 400ms |
| 19 | 5 % | 91% | 400ms |
| 22 | 10 % | 100% | 400ms |
| 20 | 0 % | 66% | 1000ms |
| 20 | 0 % | 83% | 1000ms |
| 21 | 5 % | 91% | 1000ms |

## Supplementary Note S4: White correction

All images were white-balanced to correct for the spectrum of the excitation lamp and spectral sensitivity of the camera, using the RGB values obtained from imaging a ground glass diffuser (Supplementary note S4). For best PUF performance the hues should be randomly distributed in space; if all equivalent hues clumped together the complexity of the extracted dark field micrographs would be markedly reduced. To test the spatial distribution of the hue bins, we sliced a dark field micrograph of a typical PUF into strips in both the x- and y- directions. On re-binning the image locally, we find a self-similar distribution of hues across the image (Fig. S24-S26).

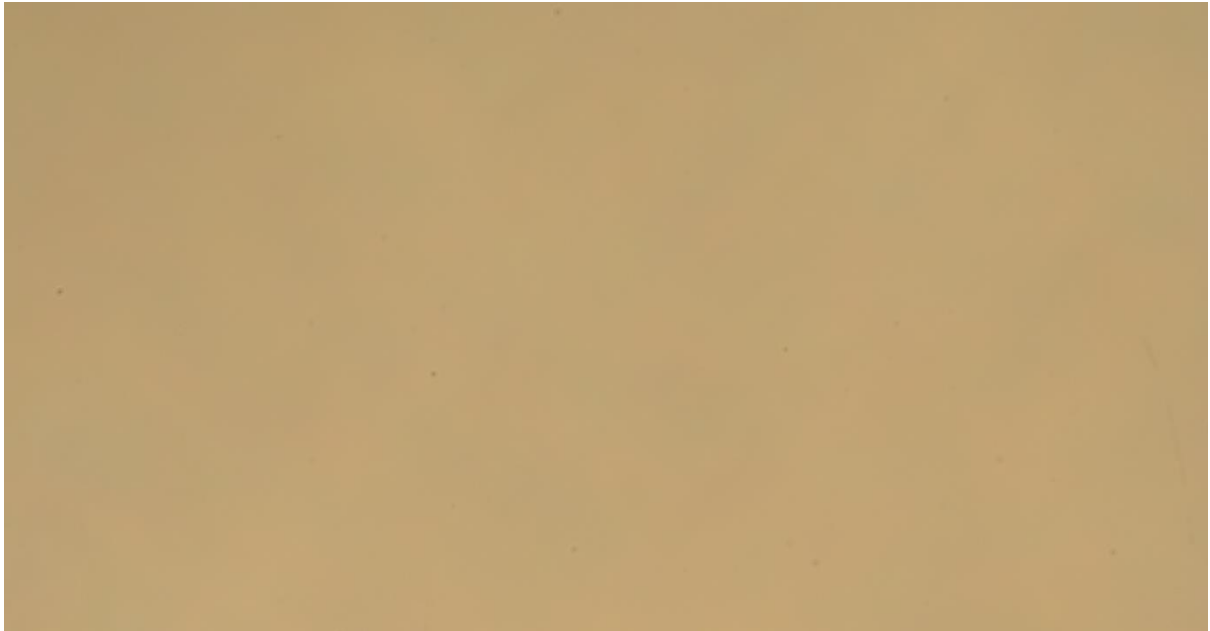

**Figure S23.** DFM image of a diffusive plate illuminated for 100 ms at 100x magnification.

We used the average RGB values (**r**: 190.9692776, **g**: 162.9511625, **b**: 116.4841115) from figure S23 and calculated the white correction constants by using the green value as baseline:

- Correction factor for green = 1
- Correction factor for red = **g/r** = 0.85
- Correction factor for blue = **g/b** = 1.40

Multiplying every pixel's RGB values by these correction factors results in white balanced images.

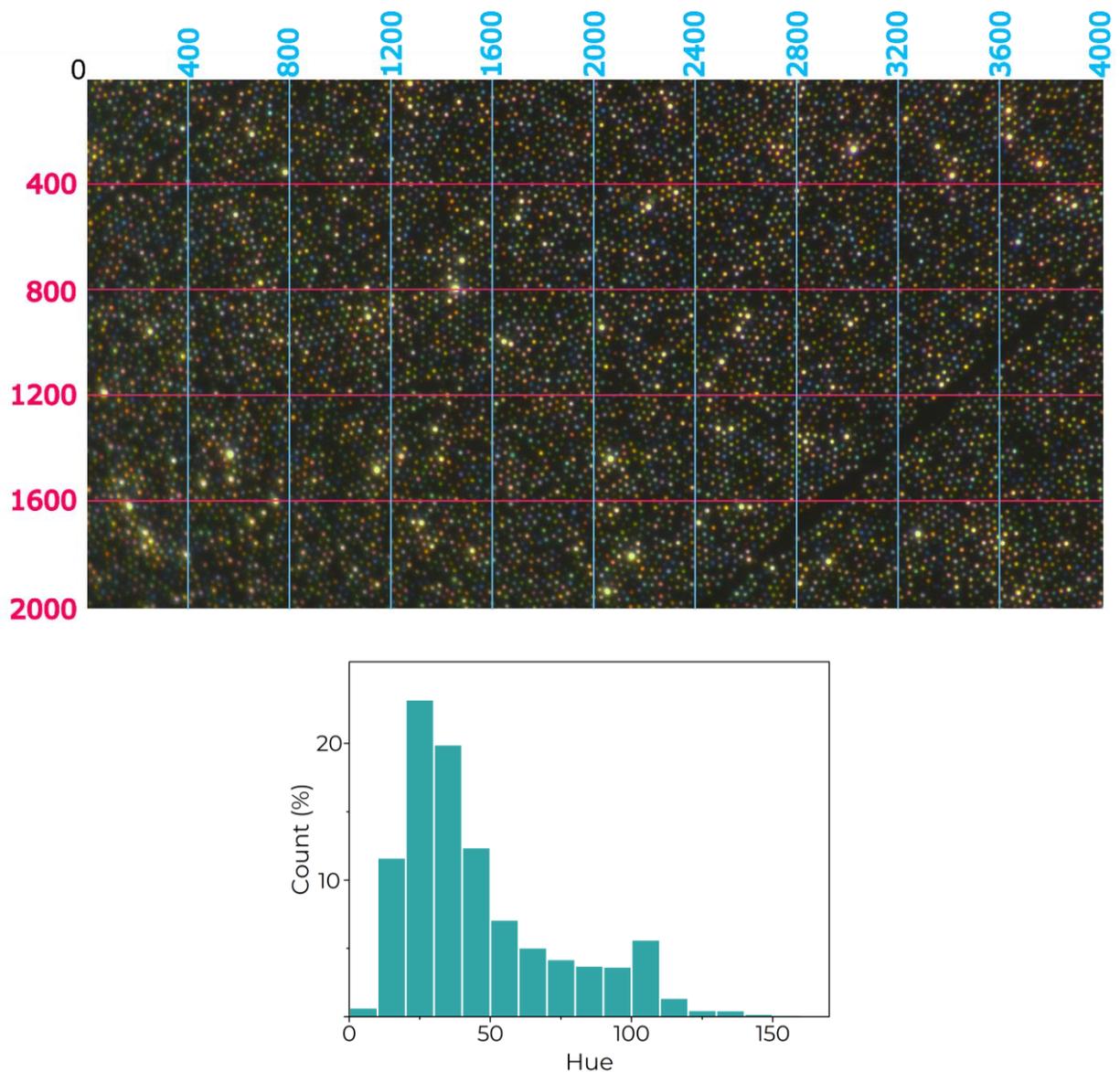

**Figure S24.** (Above) DFM image of a PUF in Fig. S20. The spacing is 1 μm. The columns analysed are shown by different colours: pink along the x-direction and blue along the y-direction. The numbers along the x- and y-directions are the pixel values. (Below) Hue histogram of the whole image.

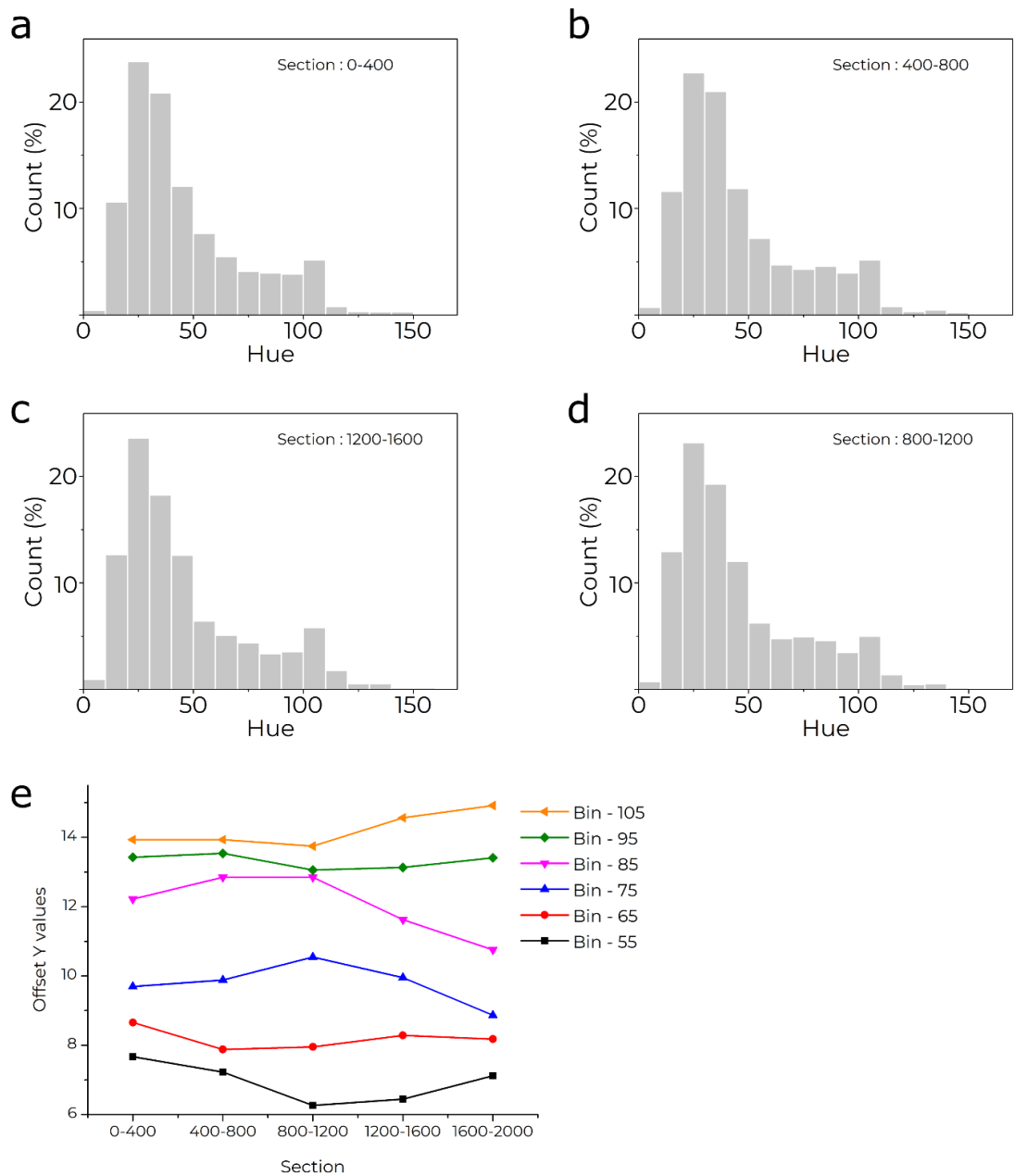

**Figure S25.** a-d) Hue histograms of four sections along the y-direction of the DFM image in Fig. S24. e) Bin count % showing small count deviations within area sections in the y-direction.

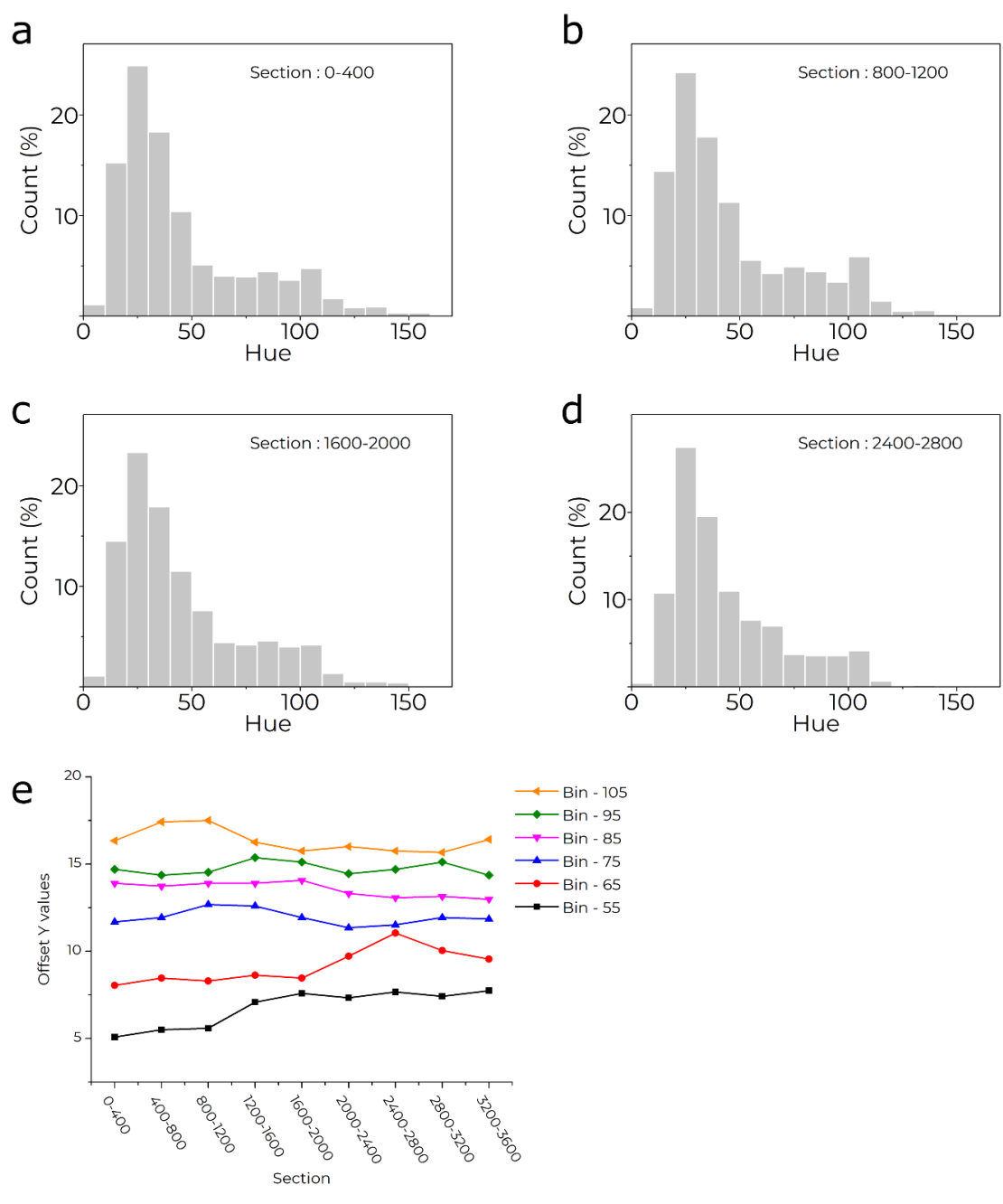

**Figure S26.** a-d) Hue histograms of four sections along the x-direction of the DFM image in Fig. S24. e) Bin count % showing small count deviations within sections in the x-direction.

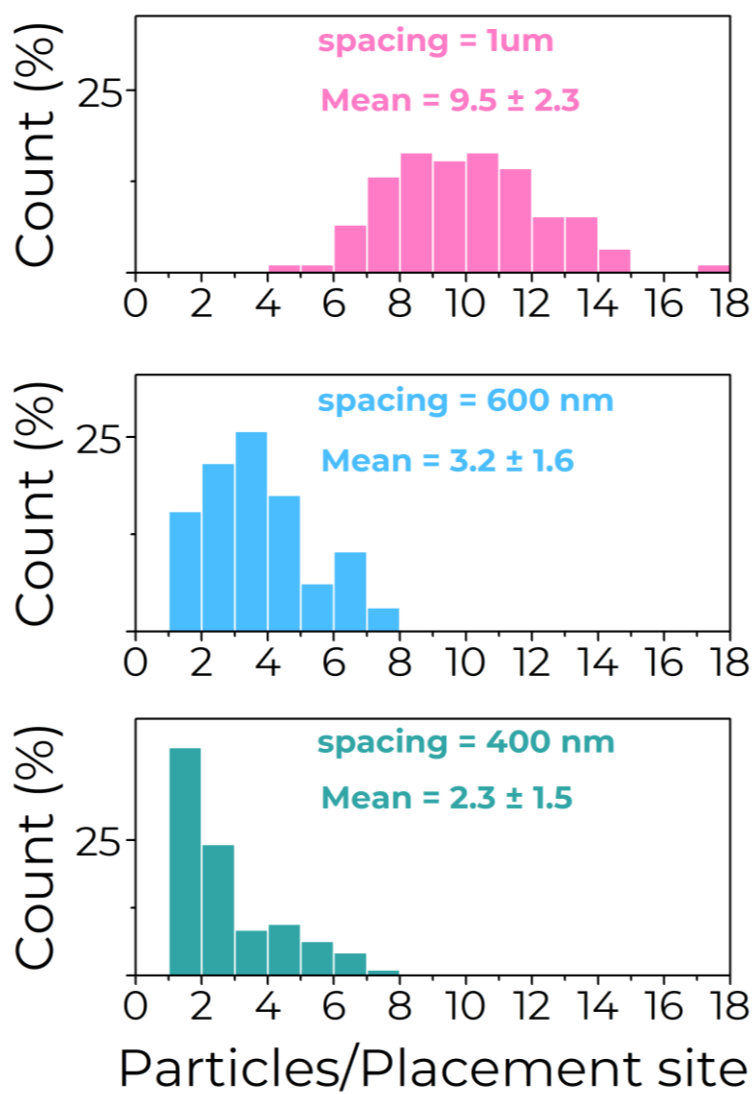

**Figure S27.** Counting analysis for number of nanoparticles occupying placement sites for PUFs made with different spacing in Fig. 2. N > 90 sites counted for each.

## Supplementary Note S5: DFM images of nanoparticles

Each variety of nanoparticle was separately suspended in water, deposited on separate plasma cleaned glass slides at O.D. ~ 0.1 and incubated for 30 s. The slides were then dried with $N_2$ flow and imaged using DFM (Fig. S28-S31). Nanorods have a tendency to stack along their long-axis, resulting in a wider variety of colours. The scattering spots are much dimmer than those in typical PUFs since individual nanoparticles scatter less light than aggregates. **No DOP or NPP was performed. No DNA origami was used.**

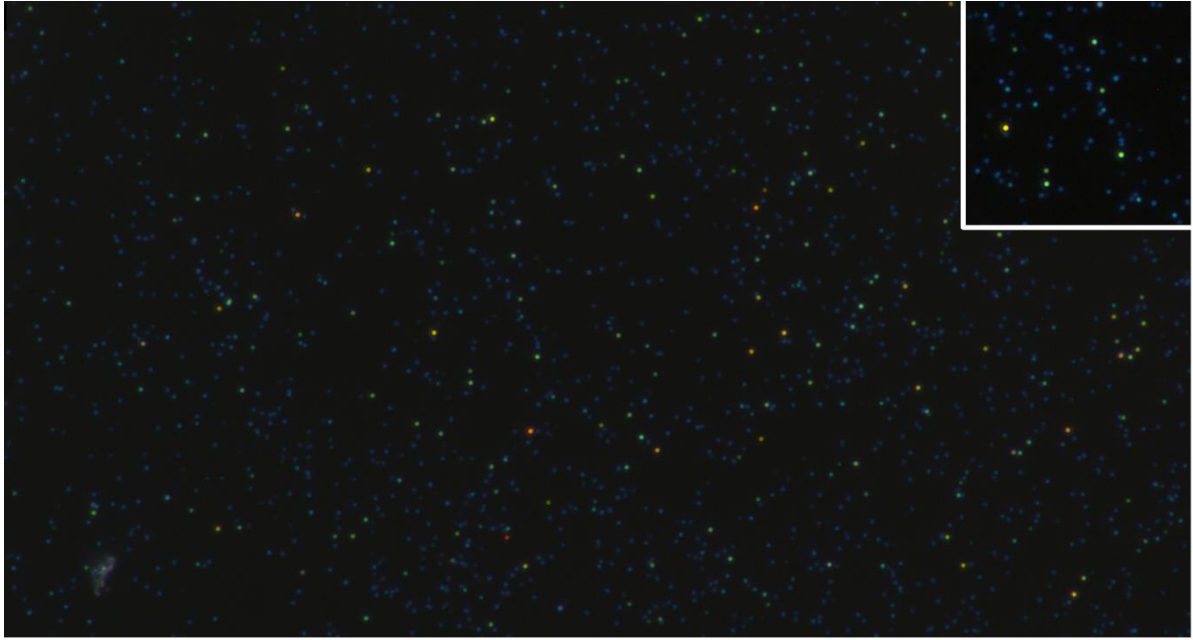

**Figure S28.** DFM images of AgNS deposited on glass slides recorded at 100x magnification. Insets: Magnified view with enhanced contrast to ease visualization.

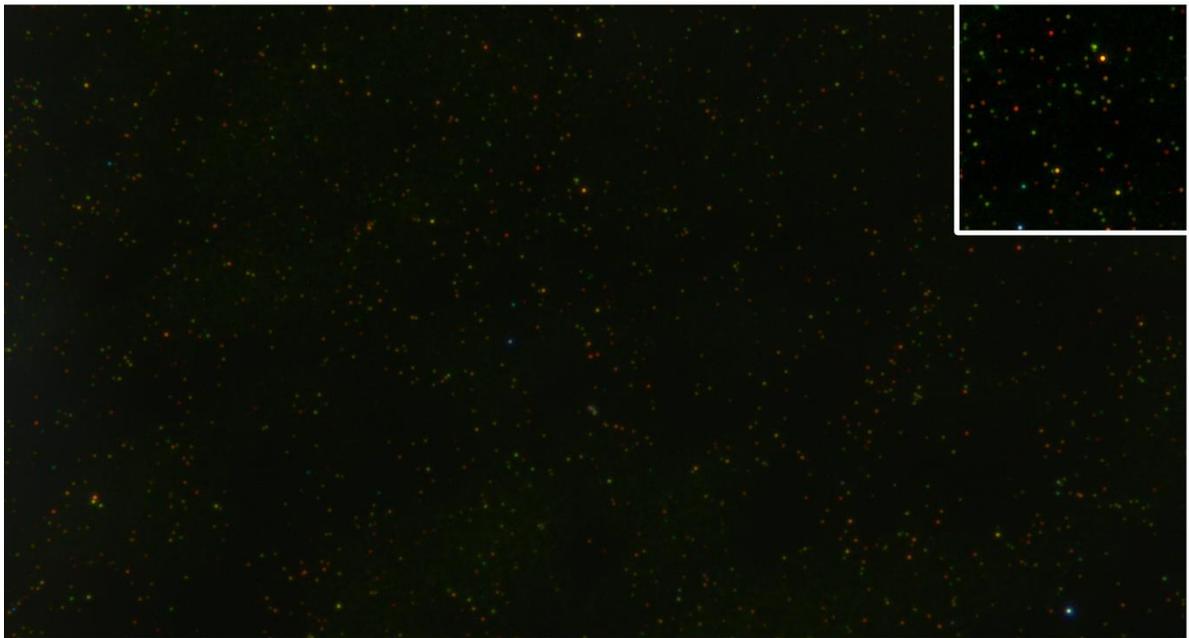

**Figure S29.** DFM images of AgNR deposited on glass slides recorded at 100x magnification. Insets: Magnified view with enhanced contrast to ease visualization.

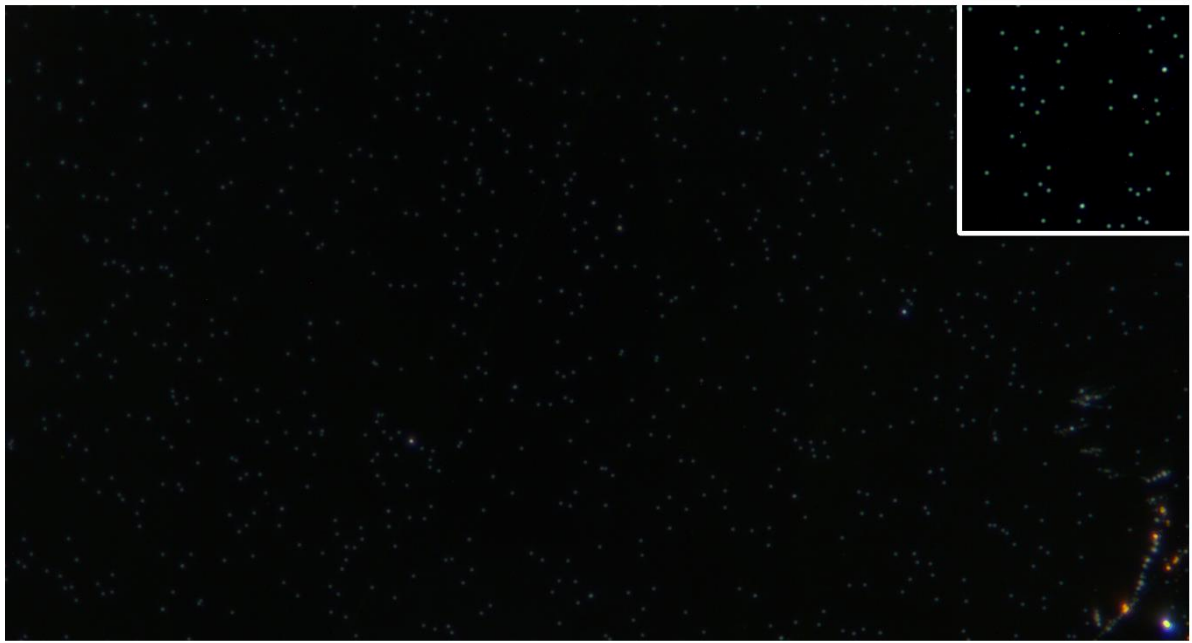

**Figure S30.** DFM images of AuNS deposited on glass slides recorded at 100x magnification. Insets: Magnified view with enhanced contrast to ease visualization.

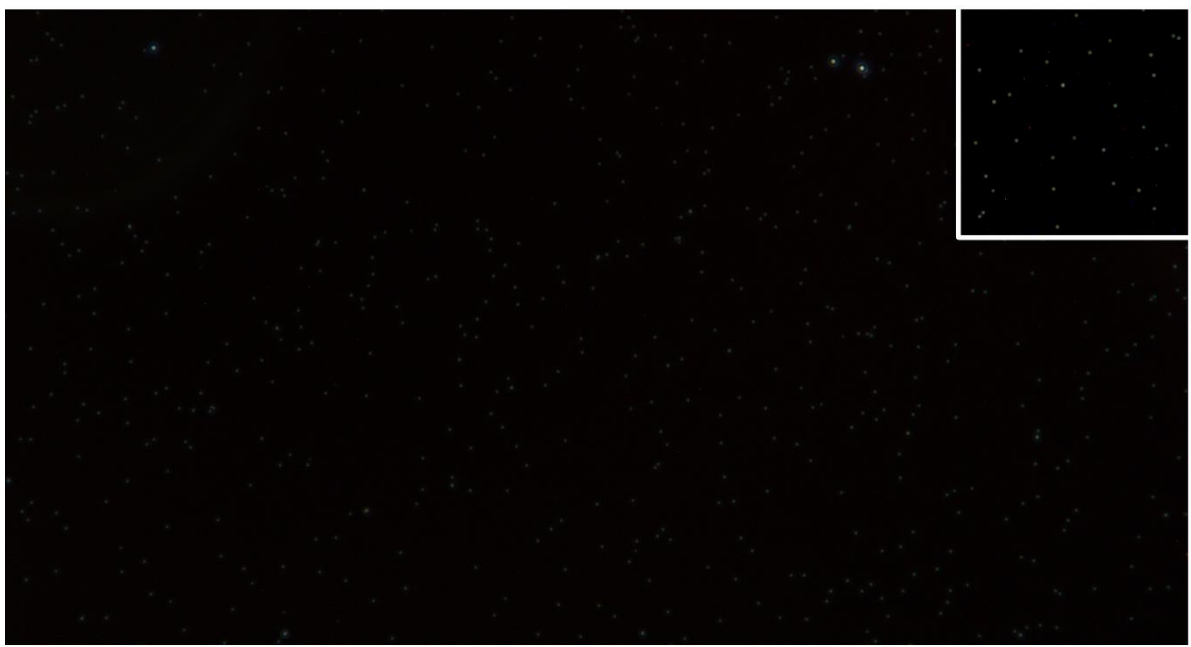

**Figure S31.** DFM images of AuNR deposited on glass slides recorded at 100x magnification. Insets: Magnified view with enhanced contrast to ease visualization.

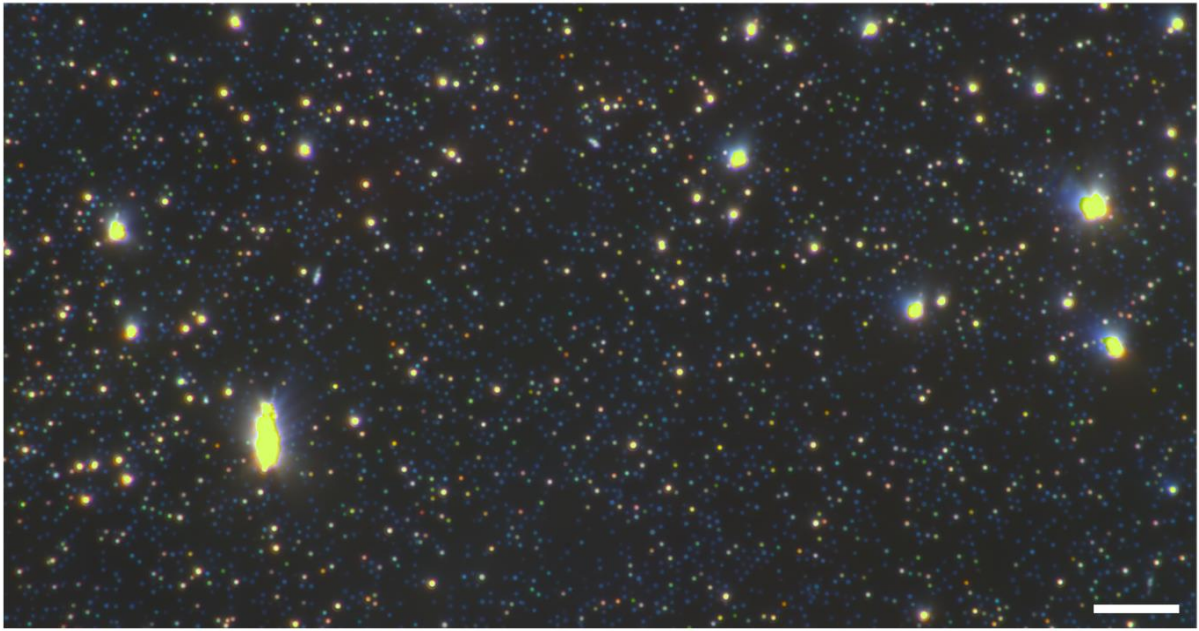

**Figure S32.** DFM image of a PUF with 1 µm spacing. Particles used – AgNS @ 0.6 OD. NPP was performed with AgNS directly **without** a DOP step. A pattern of scattering spots is visible, showing that it is possible to achieve placement of nanoparticles without DNA Origami. However, scattering spot density is low and the hue response is largely monochromatic, with a majority of the scattering spots showing blue hues similar to individual AgNS. Scale bar, 10 µm.

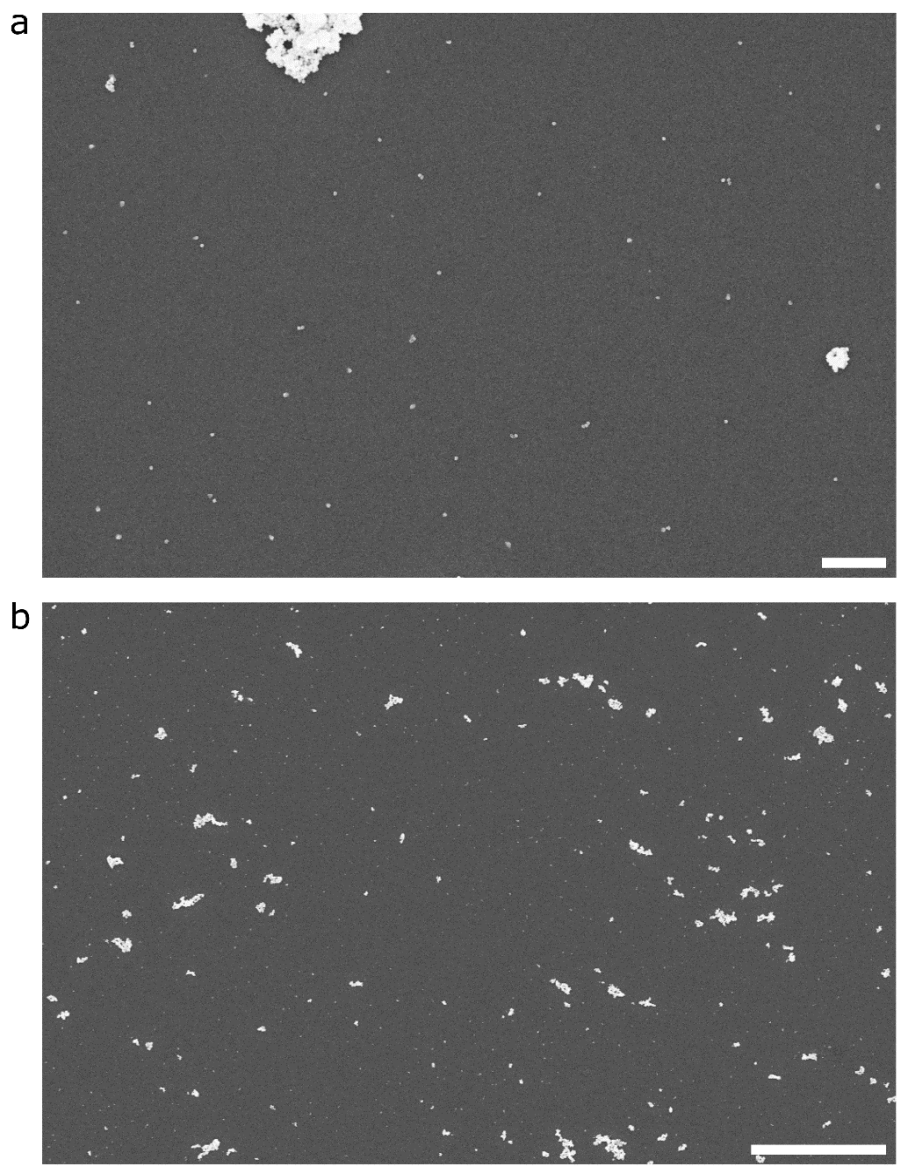

**Figure S33.** SEM images of the PUF in figure S32. Particles used – AgNS @ 0.6 OD. NPP was performed with AgNS directly **without** a DOP step. The placement consists of mostly single nanospheres at very low surface density, along with some very large aggregates, consistent with the DFM image (Fig. S32). Scale bars, (a) 1 μm, (b) 10 μm.

## Supplementary Note S6: 3-D printed DFM

The 3DFM uses linear bearings to achieve smooth movement of the objective along the steel rods. Controlled movement of the objective is realized with a threaded rod and a nut. The threaded rod and the nut are connected to the framework and the objective, respectively. The threaded rod can rotate along its long axis, screwing the nut up and down and moving the objective closer or further from the sample. An ocular focuses the image on the CCD colour-sensor of the camera module. The camera module is connected to a Raspberry Pi 3B which controls the camera module via the libcamera-still library. The colour gain values were calibrated with a white sample and set to 3.6 and 1.1 for red and blue, respectively. Therefore, the roughly 650€ prototype microscope enables imaging of our PUFs. To further reduce the cost of the microscope one could engineer a 3D-printed dark-field condenser reducing the price by 50%. Utilizing a different ocular would reduce the price by an additional 90€ to a total cost of approximately 250€.

***Table S6. Cost-table for the 3DFM.***

| Component | Cost |
|---|---|
| Condenser - LACERTA (CDF MAX) | 349.90€ |
| Objective - Kern (OBB-A1113) | 123.26€ |
| Achromatic Doublet - Thorlabs (AC254-030-A-ML) | 111.15€ |
| Camera - Raspberry Pi High Quality Camera (12MP) | 54.61€ |
| | **Total : 638.92€** |

## Supplementary Note S7: E-beam lithography nanopatterning

A *Raith eLINE* SEM was used to perform lithography on a 1 cm x 1 cm Si/SiO2 wafer with 100 nm thermal oxide (MicroChemicals). The wafer was primed with 10 ml hexamethyldisilazane (HMDS) in a 4 L desiccator. The time of priming was optimized to maintain a Si/SiO2 surface contact angle of 70–75° after HMDS deposition. The binding sites were patterned into a poly(methyl methacrylate) resist by electron-beam lithography. Then, the wafer was developed with a 1:3 solution of methyl isobutyl ketone and isopropanol. The HMDS in the developed areas was removed with $O_2$ plasma for 6 s in a plasma cleaner (Pico). The resist was stripped by ultrasonication in N-methyl pyrrolidone at 50 °C for 30 min. The substrates were briefly rinsed with isopropanol, then dried in a nitrogen stream and used immediately.

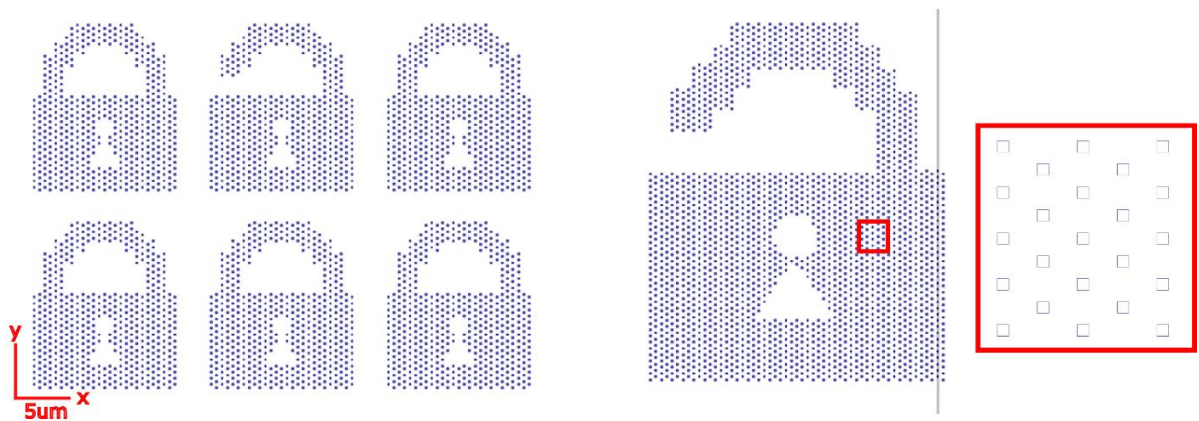

**Figure S34.** Schematic of the patterned area. Square binding sites with a side length of 150 nm are arranged in a hexagonal lattice with 600 nm period. The exposure dose was 500 uCu/cm². Inset shows a zoomed-in area of the lattice.

# References

1. González-Rubio, G. *et al.* Disconnecting Symmetry Breaking from Seeded Growth for the Reproducible Synthesis of High Quality Gold Nanorods. *ACS Nano* **13**, 4424–4435 (2019).

2. Dass, M., Kuen, L., Posnjak, G., Burger, S. & Liedl, T. Visible wavelength spectral tuning of absorption and circular dichroism of DNA-assembled Au/Ag core–shell nanorod assemblies. *Mater. Adv.* **3**, 3438–3445 (2022).

3. Shetty, R. M., Brady, S. R., Rothemund, P. W. K., Hariadi, R. F. & Gopinath, A. Bench-Top Fabrication of Single-Molecule Nanoarrays by DNA Origami Placement. *ACS Nano* **15**, 11441–11450 (2021).

4. Gopinath, A. & Rothemund, P. W. K. Optimized Assembly and Covalent Coupling of Single-Molecule DNA Origami Nanoarrays. *ACS Nano* **8**, 12030–12040 (2014).

5. Nguyen, L. *et al.* Chiral Assembly of Gold–Silver Core–Shell Plasmonic Nanorods on DNA Origami with Strong Optical Activity. *ACS Nano* **14**, 7454–7461 (2020).

6. Wei, W., Bai, F. & Fan, H. Oriented Gold Nanorod Arrays: Self-Assembly and Optoelectronic Applications. *Angewandte Chemie International Edition* **58**, 11956–11966 (2019).